\documentclass[traditabstract,longauth]{aa}
\usepackage{graphicx,amsmath,amsfonts,amssymb}
\usepackage{subfig} 
\usepackage{epsf}
\usepackage{url}
\usepackage[breaklinks, colorlinks, citecolor=blue]{hyperref}

\providecommand{\sorthelp}[1]{}

\usepackage{natbib}
\bibpunct{(}{)}{;}{a}{}{,} % to follow the A&A style

\renewcommand{\aap}{A\&A}

\renewcommand{\apjl}{ApJL}

\begin{document}

\title{A Bayesian method for point source polarization estimation}

\titlerunning{A Bayesian method for point source polarization estimation}

\author{D. Herranz\inst{1}
  \and F. Argüeso\inst{2,4}
  \and L. Toffolatti\inst{3,4}
  \and A. Manj\'on-Garc\'ia\inst{1,5}
  \and M. L\'opez-Caniego\inst{6}
  }

\institute{Instituto de F\'isica de Cantabria, CSIC-UC, Av. de Los Castros s/n, E-39005 Santander, Spain
    \and Departamento de Matem\'aticas, Universidad de Oviedo, C. Federico Garc\'ia Lorca 18, 33007 Oviedo, Spain
  \and Departamento de F\'isica, Universidad de Oviedo, C. Federico Garc\'ia Lorca 18, 33007 Oviedo, Spain
  \and Instituto Universitario de Ciencias y Tecnolog\'\i as Espaciales de Asturias (ICTEA), 
  Escuela de Ingenier\'\i a de Minas, Materiales y Energ\'\i a de Oviedo, C. Independencia 13, 33004 Oviedo, Spain
    \and Departamento de F\'isica Moderna, Universidad de Cantabria, 39005-Santander, Spain
  \and ESAC, Camino Bajo del Castillo s/n, 28692 Villafranca del Castillo, Madrid,  Spain} 

\authorrunning{Herranz et al.}

\date{}

\abstract{The estimation of the polarization $P$ of extragalactic compact sources in Cosmic Microwave Background images is a very important task in order to clean these images for cosmological purposes --as, for example, to constrain the tensor-to-scalar ratio of primordial fluctuations during inflation-- and also to obtain relevant astrophysical information about the compact sources themselves in a frequency range,  $\nu \sim 10$--$200$ GHz, where observations have only very recently started to be available. In this paper we propose a Bayesian maximum a posteriori (MAP) approach estimation scheme which incorporates prior information about the distribution of the polarization fraction of extragalactic compact sources between 1 and 100 GHz. 
We apply this Bayesian scheme to white noise simulations and to more realistic simulations that include 
CMB intensity, Galactic foregrounds and instrumental noise 
 with the characteristics of the QUIJOTE experiment Wide Survey at 11 GHz. Using these simulations, we also compare our Bayesian
 method with the frequentist Filtered Fusion method that has been already used in WMAP data and in the \emph{Planck} mission. We find that the Bayesian method allows us to decrease the threshold for a feasible estimation of $P$ to levels below $\sim 100$ mJy (as compared to $\sim 500$ mJy that was the equivalent threshold for the frequentist Filtered Fusion). We compare the bias introduced by the Bayesian method and find it to be small in absolute terms. Finally, we test the robustness of the Bayesian estimator against uncertainties in the prior and in the flux density of the sources.  We find that the Bayesian estimator is robust against moderate changes in the parameters of the prior  and almost insensitive to realistic errors in the estimated photometry of the sources. }


\maketitle

\section{Introduction} \label{sec:intro}

The polarization properties of extragalactic radio sources (ERS) --i.e., radio galaxies, radio loud quasars, blazars, etc.-- are not well constrained even at cm wavelengths, given that the total linear polarization of ERS, $P$, in general constitutes a small fraction of their total flux density, $S$. The observed value of $P$ being typically a few per cent, with only very few ERS showing a total polarization fraction, $\Pi= P/S$, as high as $\sim 10$ per cent of the total flux density
\citep[e.g.][]{Sajina11,Tucci12}.   Moreover, at shorter wavelengths, i.e. at $\lambda \leq 1$ cm,  these properties are still poorly known, due to the difficulty to properly calibrate in the radio to mm regime that afflicted the polarization experiments until a few years ago. However, the knowledge of the total and polarization fraction of ERS is rapidly improving at high radio frequencies thanks to large samples of sources mainly observed by the Australia Telescope Compact Array (ATCA) and by the Very Large Array (VLA)
\citep{sadler06,lopezcaniego09,massardi08,massardi11,massardi13,AT20G,jackson10,galluzzi17,galluzzi18}.
 More recently, and thanks to the very high sensitivity of the new detectors of the Atacama Large Millimeter Array (ALMA), 
 \cite{galluzzi19}
 could extend up to 97.5 GHz the analysis of polarization properties of ERS performed by
 \cite{galluzzi18}, by  polarimetric observations of a complete sample of 32 extragalactic radio sources. Their findings  showed that  the distribution  of the observed $\Pi$ fractions is, again, well fitted by a log-normal distribution, thus confirming previous outcomes at lower frequencies
 \citep{massardi13,galluzzi18}
 and also the predictions of 
 \cite{Tucci12}.
 The analysis of 
 \cite{galluzzi19}
 also confirmed the absence of any statistically significant trend of polarization properties of ERS with the frequency or the flux density.

Recent analyses of the ERS present in the full sky cosmic microwave background (CMB) anisotropy maps in polarization provided by the European Space Agency (ESA) Planck mission 
\citep{planck2014-a01,planck2014-a35}
also indicate typical median polarization fractions of ERS of $2 - 3$\% at frequencies as high as 300 GHz
\citep{bonavera17a,bonavera17b,trombetti18}. 
Therefore, an accurate characterization of polarization properties of ERS as well as their efficient detection and subtraction from CMB maps is especially crucial for measuring the primordial CMB B-mode polarization down to values of the tensor to scalar ratios $ r \sim 0.001$, that could be achievable by future space probes
(i.e. \citealp[LiteBird:][]{LiteBird};  \citealp[COrE:][]{delabrouille18}). We remind that the simulations by 
\cite{remazeilles}
 have shown that, at these low values of $r$, unresolved polarized ERS can probably be the dominant foreground at multipoles $\ell > 50$ in the power spectrum of the CMB anisotropy. These results have been confirmed by 
 \cite{puglisi18},
 by exploiting the state-of-the-art data sets on polarized point sources over the 1.4--217 GHz frequency range. 

In addition to the essential information that the polarization of ERS provides about the structure and evolution of extragalactic baryonic matter at low to intermediate redshifts, the study of this polarized radiation is paramount for cosmology, and in particular for Cosmic Microwave Background (CMB) science.  ERS detection and subtraction is  a fundamental part of the component separation process necessary to achieve the science goals set for the next generation of CMB experiments. In particular, ERS would significantly affect the estimation of the CMB polarization angular power spectra and, therefore, limit the ability of CMB experiments to constrain cosmological parameters such as the tensor-to-scalar ratio $r$ of primordial perturbations during inflation. 
ERS could become an important obstacle for the detection of the Primordial Gravitational Wave Background (PGWB) for low values of $r$ \citep{B-mode_detectability_Tucci,puglisi18, Trombetti}  due 
to both the additional noise they constitute in themselves, and the reduction in delensing power they cause by degrading lensing potential reconstructions \citep[see, e.g.,][]{pst}. Therefore, during recent years the interest in the development of signal processing techniques specifically tailored for the detection and characterization of ERS in CMB images has been growing in the literature.

Signal processing techniques for the detection of polarized ERS must take into account the spinorial nature of electromagnetic waves.  The signal can be described by not only one but as many as four independent components, one for the total intensity of the radiation field and three for its polarization state. It is convenient to use the Stokes' parameters $S, Q, U, V$ ($S$ for total intensity\footnote{The usual notation for this Stokes parameter is $I$. However, in this work we have changed the notation in order to avoid confusion between the intrinsic intensity of a source, that we will call later in this paper $S_0$, and the modified Bessel function of zero order, $I_0$, that appears in several equations in Section~\ref{sec:method}.}
in terms of flux density, see, e.g., \cite{galluzzi19},
$Q$ and $U$ for linear polarization and $V$ for circular polarization), but other representations are also possible.  The Stokes' $V$ parameter is not usually considered, since Thomson scattering does not induce circular polarization in the CMB. Circular polarization mechanisms in active galaxies have been described in the literature \citep[see for example][]{rayner}, but they are nonetheless considered to be sub-dominant in comparison to linear polarization mechanisms. Therefore, in this paper we will consider, as it is customary\footnote{Foregrounds can produce circular polarization under some circumstances, and it has been observed in a few extragalactic sources. The value of $V$ is typically much lower than the other Stokes' parameters. As it will be explained in Section~\ref{sec:method}, the existence of sources with non-zero circular polarization would not affect our estimations of the $Q$ and $U$ Stokes parameters. Of course, if there was a significant $V$ term, neglecting it would lead us to miss a part of the polarization $P$. However, our method can be easily adapted to work with a third component in the form of an additional image --corresponding to the $V$ Stokes’ parameter-- if necessary.}, 
$V=0$. Then the signal processing of polarized ERS must deal with three independent quantities, two of them having the mathematical structure of a spinor field. 

The $S$, $Q$, and $U$ signals (or, alternatively, $S$, $E$, and $B$, or any other set of three quantities obtained from the Stokes' parameters) can be treated separately as independent images to which any of the standard compact component separation techniques could be applied. The main difference with respect to the classical setting is that, unlike the total flux density $S$, which is always non-negative, $Q$ and $U$ can be either positive, negative or zero. From a physical point of view, however, it makes more sense to process the polarization data jointly \citep[see][for a review on the topic]{review}. In particular,  the  \textit{total polarization}  of a source $P=\sqrt{Q^2+U^2}$ and its \textit{polarization fraction} $\Pi = P/S$ are directly related to the physical processes occurring along the path of photons from the ERS to Earth, while $Q$ and $U$ are frame-dependent quantities lacking in physical meaning on themselves. 

The main two problems arising when dealing with  $P$ are the typically low signal-to-noise ratio of the polarization signal coming from ERS and the non-Gaussian distribution of its noise statistics. Regarding the former, as mentioned above the typical polarization fractions of ERS at frequencies below $\sim 10$ GHz are at most $10 \%$. This means that only a few ERS are bright enough to be detected in polarization with present-day technology. A standard procedure to avoid false detections in polarization is to \emph{detect} sources in total intensity and then to try to estimate their polarization properties in a non-blind way\footnote{That is, focusing efforts on the precise position of the source once it has been detected in intensity, i.e. the non-blindness is only related to the positions of targets, not to any other quantity.}. We will follow this approach in this paper. Regarding the latter problem, assuming that the $Q$ and $U$ noises are Gaussian-distributed, $P$ will have a non-Gaussian Rice distribution \citep{Rice}. Rician distribution has strictly non-negative support and heavy tails, which a) biases the estimation of the polarization of the sources and b) disrupts the intuitive interpretation of signal-to-noise in terms of $\sigma$ thresholds that is used virtually everywhere else in radio Astronomy.  \cite{simmons&stewart} discussed four estimators which attempted to correct for biasing in the degree of linear polarization in the presence of low signal-to-noise ratios. More recently, \cite{argueso09}  studied the problem in the context of CMB astronomy and developed two methods for the detection/estimation of ERS in polarization data: one that applies the Neyman-Pearson lemma to the Rice distribution, the Neyman-Pearson filter (NPF), and another based on pre-filtering before fusion of $Q$ and $U$ to obtain $P$, the filtered fusion (FF) method. That work found that under typical CMB-experiment settings
the FF outperforms the NPF both in terms of
computational simplicity and accuracy, especially for low fluxes.
\cite{lopezcaniego09} applied the FF to the WMAP five-year data. The same method has been used to study the polarization of the \emph{Planck} Second Catalogue of Compact Sources \citep[PCCS2,][]{planck2014-a35} and of the QUIJOTE experiment Wide Survey Source Catalogue \citep{quijote_PS}. Alternatively, a novel method for the estimation of the polarization intensity and angle of compact sources in the $E$ and $B$ modes of polarization based on steerable wavelets has been recently proposed by \cite{patricia20}.

All the previously mentioned methods attempt to estimate the ERS polarization by minimizing as much as possible the impact of noise and Galactic and extragalactic foregrounds on the observed signal. The \emph{expected} value of the polarization does not intervene in the estimation process. In other words, no \emph{a priori} information is used in the estimation. Until very recently, this has been the most sensible choice, as the polarization properties of extragalactic sources were virtually unknown at microwave frequencies. However, as recent experiments and facilities such as the ALMA, \emph{Planck} and the upgraded versions of ATCA and VLA  start shedding light on the $\lambda \leq 1$ cm polarized sky, the possibility of adding physical priors to our signal processing techniques is gradually opening. In this paper, we propose a Bayesian maximum \emph{a posteriori} (MAP) method for the estimation of the polarization properties of point sources.

The structure of this paper is as follows. In section~\ref{sec:method} we review the current observational evidence to construct physical priors on the polarization fraction of ERS and incorporate that information into two possible MAP estimators of the polarization of a compact source of known flux density $S$. These MAP estimators take a form analogous to the Neyman-Pearson filter and Filtered Fusion by \cite{argueso09}, respectively, plus additional terms that contain the \emph{a priori} physical information of the probability distribution function of $P$ for ERS. We call these two methods Bayesian Rice and Bayesian Filtered Fusion, respectively. The Bayesian Filtered Fusion is easily applicable for both white and colour noise. For this reason, and because in \cite{argueso09} it was shown that the FF outperforms the NPF, we focus the rest of the paper on the Bayesian Filtered Fusion. 
In section~\ref{sec:simulations} we describe the simulations we have used to test the Bayesian Filtered Fusion method. We first make simplistic simulations containing just white noise in $Q$ and $U$ and then we upgrade to realistic simulations with polarized Galactic foregrounds and CMB emissions. In both cases, we use angular resolution, pixel scale and noise levels similar to the upcoming QUIJOTE experiment Wide Survey data at 11 GHz \citep{mfiwidesurvey}. The results of applying the Bayesian Filtered Fusion to our simulations are discussed in section~\ref{sec:results}, where we also make a brief discussion about the robustness of the method against uncertainties on the priors and the determination of the total flux density of the sources. Finally, we draw our conclusions in section~\ref{sec:conclusions}.

\section{Method}  \label{sec:method}

When we try to detect or estimate the polarization $P_0$ of a compact source embedded in Gaussian noise --what can be a good approximation when dealing with sources present in CMB maps--, and we consider the measured polarization $P =  \sqrt{Q^2 + U^2}$ with similar Gaussian noise dispersions in $Q$ and $U$, i.e. $\sigma_Q = \sigma_U = \sigma$, the distribution of 
$P$ given $P_0$ follows the Rice distribution
\begin{equation} \label{eq:probP_gP0}
f \left( P | P_{ 0 } \right) = P / \sigma ^ { 2 } \exp \left[ - \left( P ^ { 2 } + P_{ 0 }^{ 2 } \right) / 2 \sigma ^ { 2 } \right] I_{ 0 } \left( P P_{ 0 } / \sigma ^ { 2 } \right) 
\end{equation}
\noindent
This is the conditional probability distribution of $P$, measured polarization, given $P_0$, source polarization, with $I_0$ the modified Bessel function of zero order \citep{Rice}. This distribution has been used to obtain suitable estimators of $P_0$ \citep{simmons&stewart,argueso09,lopezcaniego09,review}. However, no previous knowledge about $P_0$ is assumed in these papers and this information, if available, could be very useful after being incorporated in a Bayesian scheme. Recently, relevant data about the
distribution of the polarization fraction at different frequencies, $\Pi = P_0/S_0$ , with $S_0$
the source flux density, have been presented \citep{massardi13,galluzzi17,galluzzi19}. According to these authors, this distribution can be represented by a log-normal probability density function (pdf)
\begin{equation} \label{eq:lognormal}
g ( \Pi ) = \frac{ 1 } { \sqrt{ 2 \pi } \sigma_{ \Pi } \Pi } \exp \left[ - ( \log ( \Pi ) - \mu )^{ 2 } / 2 \sigma_{ \Pi }^{ 2 } \right]
\end{equation}
\noindent
where $\mu = \log \Pi_{med}$ --with $\Pi_{med}$ the median polarization fraction-- and $\sigma_{\Pi}$ are easily obtained from $\langle \Pi \rangle$ and $\langle \Pi^2 \rangle$ \citep{Crow}. Since $P_0 = \Pi S_0$, if we assume that the value of $S_0$ is known, then the distribution of $P_0$ is readily calculated
\begin{equation} \label{eq:probP0}
g \left( P_{ 0 } \right) = \frac{ 1 } { \sqrt{ 2 \pi } \sigma_{ \Pi } P_{ 0 } } \exp \left[ - \left( \log P_{ 0 } - \mu_{ 1 } \right)^{ 2 } / 2 \sigma_{ \Pi }^{ 2 } \right]
\end{equation}
\noindent
with $\mu_1 = \mu + \log\left(S_0\right) = \log\left(S_0 \Pi_{med}\right)$. 
This assumption is safe, since estimating $S_0$ is much simpler than estimating $P_0$ and point sources for which the polarized emission is detectable tend to have high flux densities. Therefore,  we will work, in general, with non-blind detection in polarization. The knowledge of $S_0$ allows us to write the joint probability distribution of $P$ and $P_0$, $h(P,P_0) = f(P|P_0) g(P_0)$, and by using Bayes’ theorem
\begin{equation} \label{eq:probP0gP}
f \left( P_{ 0 } | P \right) = \frac{ h\left( P , P_{ 0 } \right) } { g ( P ) }
\end{equation}
\noindent
with $g ( P ) = \int h \left( P , P_{ 0 } \right) d P_{ 0 }$. Finally, we obtain, by substituting (\ref{eq:probP_gP0}) and (\ref{eq:probP0}) in (\ref{eq:probP0gP}),
\begin{equation}
f \left( P_{ 0 } | P \right) = \frac{ e^{ - P_{ 0 }^{ 2 } / 2 \sigma ^ { 2 }} I_{ 0 } \left[\frac{ P P_{ 0 }}{  \sigma ^ { 2 }} \right] \exp \left[ - \frac{\left( \log P_{ 0 } - \mu_{ 1 } \right)^2 }{ 2 \sigma_{\Pi } ^2} \right] \frac{ 1 }{P_0} } { \int{ e^{ - P_{ 0 }^{ 2 } / 2 \sigma ^ { 2 }} I_{ 0 } \left[\frac{ P P_{ 0 }}{  \sigma ^ { 2 }} \right] \exp \left[ - \frac{\left( \log P_{ 0 } - \mu_{ 1 } \right)^2 }{ 2 \sigma_{\Pi } ^2} \right] \frac{ d P_0 }{P_0}} } .
\end{equation}
\noindent
The integral in the denominator is just a normalization. We have found the distribution of $P_0$ given $P$, the posterior distribution, simply by assuming Gaussian noise in $Q$ and $U$ with the same dispersion and a log-normal pdf for $\Pi$ (prior distribution).
Everything has been calculated for a source located at a central pixel and without taking into account any information about the beam and the data in a certain patch around the source. If we consider a polarized source at the central pixel of an $n$-pixel patch, a beam with profile $\tau(\mathbf{x})$ and values $P_i$ for the polarization measured at each pixel $i = 1, \ldots, n$, we can write the following expression for the conditional pdf of $P_0$ given the values $P_i$ at the different pixels:
\begin{eqnarray}
f \left( P_{ 0 } | P_{ i } \right) & \propto & \exp \left[ - \frac{\left( \log P_{ 0 } - \mu_{ 1 } \right)^{ 2 } }{ 2 \sigma_{ \Pi }^{ 2 }} \right] \frac{ 1 } { P_{ 0 } } \times \nonumber \\ 
& & \prod_ { i } \exp \left[ - P_{ 0 }^{ 2 } \frac{ \tau_{ i }^{ 2 } } { 2 \sigma_{ i }^{ 2 } } \right] I_{ 0 } \left[ P_{ 0 } \frac{ P_{ i } \tau_{ i } } { \sigma_{ i }^{ 2 } } \right]
\end{eqnarray}
\noindent
Here $\Pi_i$ is the product symbol, $\tau_i$ is the profile at each pixel and $\sigma_i$ the noise dispersion (it could be different from pixel to pixel). We take the natural logarithm of the right-hand side and change sign, this is minus the log-posterior of the distribution, save constant terms. In this way, we obtain a simplified expression that we will later
minimize to find the estimator that makes the posterior distribution maximum
\begin{eqnarray} \label{eq:PBayes}
- \log f \left( P_{ 0 } | P_{ i } \right) & = & \frac{\left( \log P_{ 0 } - \mu_{ 1 } \right)^{ 2 } }{ 2 \sigma_{ \Pi }^{ 2 }} + \log \left( P_{ 0 } \right) + P_{ 0 }^{ 2 } z \nonumber \\
&  - & \sum_{ i } \log I_{ 0 } \left[ P_{ 0 } \, y_{ i } \right] + K
\end{eqnarray}
\noindent
with
\begin{equation}
 z = \sum_{ i } \frac{ \tau_{ i }^{ 2 } } { 2 \sigma_{ i }^{ 2 } }  
 \end{equation}
 \noindent
and
\begin{equation}
   y_{ i } =  \frac{ P_{ i } \tau_{ i } } { \sigma_{ i }^{ 2 } }  ,
\end{equation}
\noindent 
and where $K$ is a constant term that encloses the
proportionality terms not directly included in (\ref{eq:PBayes}).
If we differenciate with respect to $P$ and equate to zero, the estimator $\hat{P}_0$ will satisfy
\begin{equation}
\frac{ \log \hat { P }_{ 0 } - \mu_{ 1 }  } { \hat { P }_{ 0 } \sigma_{ \Pi }^{ 2 }} + \frac{ 1 } { \hat { P }_{ 0 } } + 2 \hat { P }_{ 0 } z - \sum_{ i } \frac{ I_{ 1 } \left[ \hat { P }_{ 0 } y_{ i } \right] } { I_{ 0 } \left[ \hat { P }_{ 0 } y_{ i } \right] } \, y_{ i } = 0
\end{equation}
\noindent
with $I_1$ the modified Bessel function of order one. On the other hand,  in \cite{argueso09} a method called filtered fusion (FF) was shown to perform better than the one derived from the Rice distribution. The FF calculates the square root of the sum of the squares of the maps in Q and U to which a matched filter has been previously applied. This method is just a maximization of the conditional probability of the data $Q_i$, $U_i$ given the source polarization $Q_0$, $U_0$, assuming that the noise is Gaussian with zero mean and independent for each pixel
\begin{equation}
f(Q_i,U_i|Q_0,U_0)=\prod_ { i }  \exp \left[ \frac{ \left( Q_{ i } - Q_{ 0 } \tau_{ i } \right)^{ 2 } } { 2 \sigma_{Q_ i }^{ 2 } } +\,
 \frac{ \left( U_{ i } - U_{ 0 } \tau_{ i } \right)^{ 2 } } { 2 \sigma_{U_ i }^{ 2 } } \right]
\end{equation}

If we also assume that $P_0$ follows a log-normal pdf and the polarization angle distribution is uniform, we can combine the previous formula with the prior distribution of $Q_0,U_0$ and write, by applying Bayes’ theorem, minus the log-posterior of $Q_0$, $U_0$ given the data $Q_i$, $U_i$ 
\begin{eqnarray} \label{eq:estimator}
- \log f \left( Q_{ 0 } , U_{ 0 } | Q_{ i } , U_{ i } \right) & = & \frac{\left( \log \sqrt{ Q_{ 0 }^{ 2 } + U_{ 0 }^{ 2 } } - \mu_{ 1 } \right)^{ 2 } }{ 2 \sigma_{ \Pi }^{ 2 }} \nonumber \\
&  + & \log \left( Q_{ 0 }^{ 2 } + U_{ 0 }^{ 2 } \right) \nonumber \\
& + & \sum_{ i } \frac{ \left( Q_{ i } - Q_{ 0 } \tau_{ i } \right)^{ 2 } } { 2 \sigma_{ Q_i }^{ 2 } } \nonumber \\
& + & \sum_{ i } \frac{ \left( U_{ i } - U_{ 0 } \tau_{ i } \right)^{ 2 } } { 2 \sigma_{ U_i }^{ 2 } } + K.
\end{eqnarray} 
\noindent
 This expression could be very easily
generalized, allowing even the treatment of correlations between the noise in different pixels. In that case, (\ref{eq:estimator}) can be written
\begin{eqnarray} \label{eq:estimatorb}
& - & \log f \left( Q_{ 0 } , U_{ 0 } | Q_{ i } , U_{ i } \right) = \nonumber \\
 & &    \frac{\left( \log \sqrt{ Q_{ 0 }^{ 2 } + U_{ 0 }^{ 2 } } - \mu_{ 1 } \right)^{ 2 } }{ 2 \sigma_{ \Pi }^{ 2 }} \nonumber \\
&  & + \log \left( Q_{ 0 }^{ 2 } + U_{ 0 }^{ 2 } \right) \nonumber \\
& & +  \frac{1}{2} \sum_{\mathbf{k}} \left( Q_{\mathbf{k}} - Q_0 \tau_{\mathbf{k}}  \right)^t \mathbf{P}^{-1}_{Q,\mathbf{k}}  \left( Q_{\mathbf{k}} - Q_0 \tau_{\mathbf{k}}  \right) \nonumber \\
& & +  \frac{1}{2} \sum_{\mathbf{k}} \left( U_{\mathbf{k}} - U_0 \tau_{\mathbf{k}}  \right)^t \mathbf{P}^{-1}_{U,\mathbf{k}}  \left( U_{\mathbf{k}} - U_0 \tau_{\mathbf{k}}  \right) \nonumber \\
& & +  K,
\end{eqnarray} 
\noindent
where the subindex $\mathbf{k}$ refers to the Fourier wave vector (or, for the case of spherical data, the appropriate spherical harmonic) and $P_{x,\mathbf{k}}$ is the power spectrum (or angular power spectrum) of the noise for the Stokes parameter $x$. This formula is expressed in Fourier space for the sake of computational efficiency, but could be also expressed in real space by means of the correlation matrix of the noise. In (\ref{eq:estimatorb}) it is immediate to recognize that the third and fourth terms in the right side of the equation are analogous to the matched filter on the $Q$ and $U$ maps, that in turn are the solution of a Maximum Likelihood estimator (MLE). The first term adds the prior information, whereas the second term acts as a penalty for large values of the estimated  polarization, and finally the last term is a constant that is irrelevant for the solution.

Now we can obtain the estimators $\hat{Q}_0$ and $\hat{U}_0$
 that minimize the previous expression. 
Finally, we find 
\begin{equation}
\hat{P}_0 = \sqrt{\hat{Q}^2_{0} + \hat{U}^2_{0}} 
\end{equation}
\noindent
as our estimator of $P_0$.
In all these formulas we have not considered the effect of the circular polarization, $V$.  As commented in the introduction, this effect is very small and can be, in general, neglected.
At any rate, the generalization of the Rice and FF methods to include circular polarization has been presented in \cite{argueso11}. However, in order to extend our Bayesian methods to this case, we would have to use a prior distribution for $V$ which is not known yet.

To sum up the previous paragraphs: we have presented four possible estimators of $P_0$, the old ones are the Rice method and the FF \citep{argueso09,lopezcaniego09,review,planck2014-a35}, and the new ones, obtained by minimizing the right-hand sides of (\ref{eq:PBayes}) and (\ref{eq:estimator}) or (\ref{eq:estimatorb}), that we will call
 Bayesian Rice method and Bayesian FF, respectively. These new methods incorporate in a natural way our information about the source polarization distribution. The  FF and Rice methods are implemented by minimizing (\ref{eq:PBayes}) and (\ref{eq:estimator})-(\ref{eq:estimatorb}) without the first two terms, which come from the Bayesian prior.

\section{Simulations} \label{sec:simulations}

\subsection{White noise}  \label{sec:sim_white}

As a first test bed to assess  the performance of
the Bayesian techniques, 
we have run 10000 simulations using only  white noise as background for  simulated sources. Table~\ref{tb:params} shows the simulation parameters for these simulations; 
the pixel size, beam FWHM and white noise rms emulate those of the QUIJOTE
 \citep[Q-U-I JOint TEnerife,][]{QUIJOTE10,QUIJOTE12,QUIJOTE15,QUIJOTE16} experiment Wide Survey at 11 GHz \citep{mfiwidesurvey}. We will use the same simulation parameters for the full sky simulations to be discussed in Section~\ref{sec:sim_sky}.

 For these white noise simulations we directly create flat images with uncorrelated Gaussian noise and inject 
 at the center a point source with  he FWHM listed in Table~\ref{tb:params}, a given flux density $S_0$ and polarization fraction $\Pi$ randomly drawn from the log-normal distribution (\ref{eq:lognormal}) with the mean and standard deviation values $\langle \Pi \rangle$  and  $\sigma_{\Pi}$  described in Table~\ref{tb:params}.  We simulate intensities $S_0$ in ten logarithmically spaced values between 0.1 and 100 Jy. In this way we get a sample of sources from moderately faint to extra bright (and, since the polarization fraction 
follows distribution (\ref{eq:lognormal}) with average $\langle \Pi \rangle = 0.02$, values of $P$ from below 1 mJy to a few tens of Janskys).

\subsection{Full sky simulations}  \label{sec:sim_sky}

\begin{figure*}[h]
  \centering
   \includegraphics[width=\textwidth]{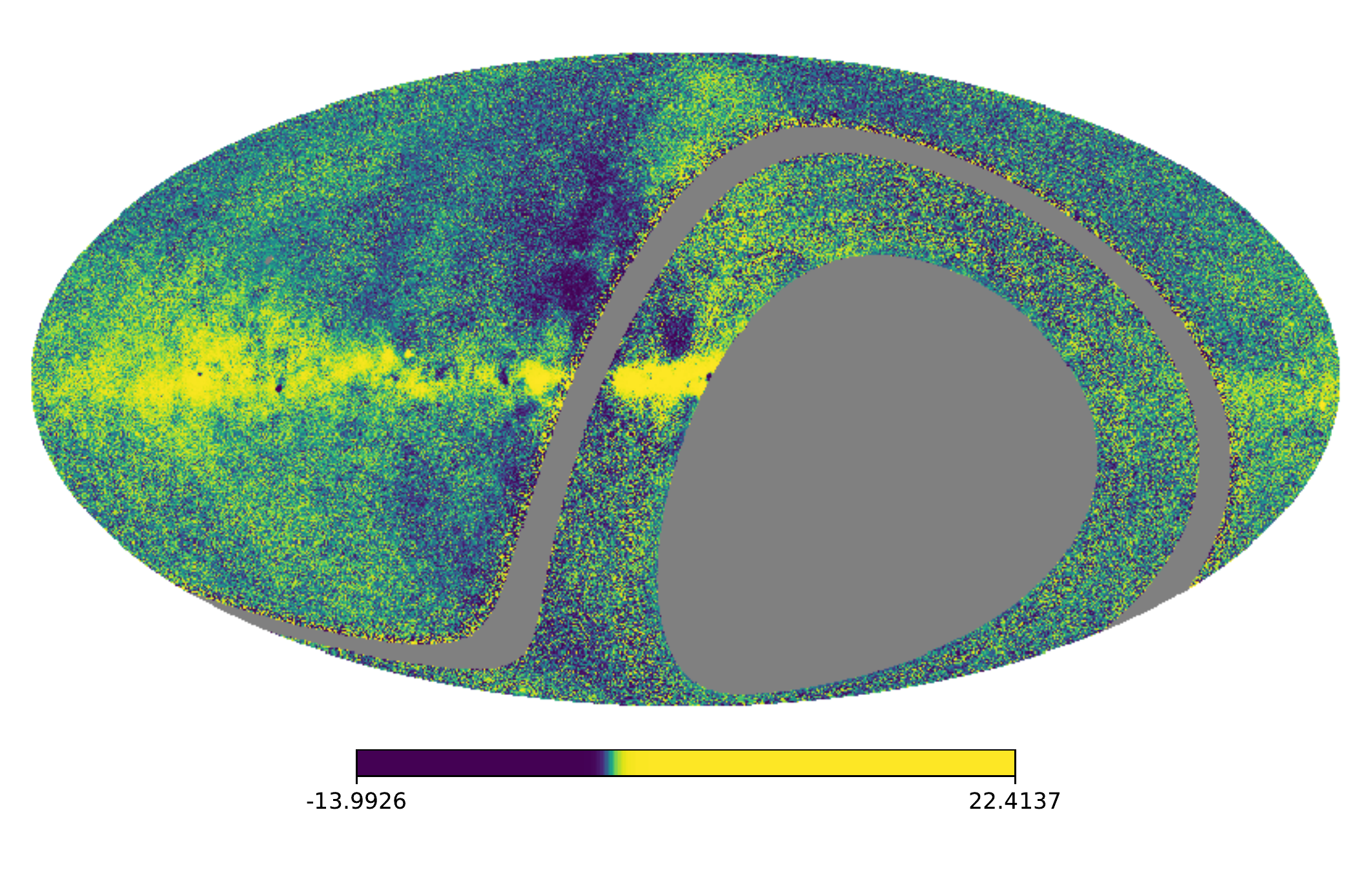}
   \caption{$Q$-Stokes simulated QUIJOTE Wide Survey sky at 11 GHz. The false colour bar is expressed in Jy.}  \label{fig:PSM_Q}
\end{figure*}

\begin{figure*}[h]
  \centering
   \includegraphics[width=\textwidth]{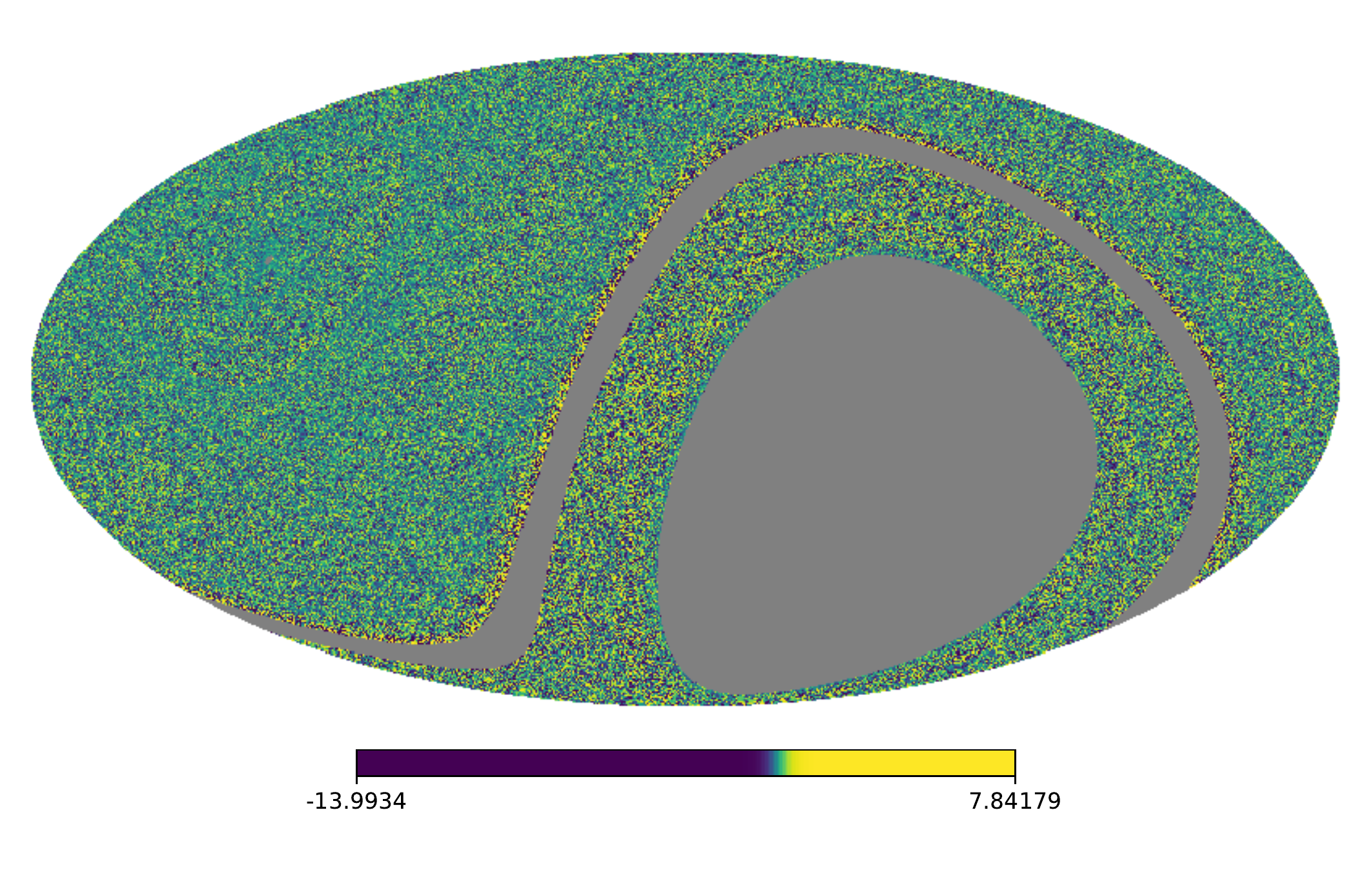}
   \caption{$Q$-Stokes simulated QUIJOTE Wide Survey instrumental noise at 11 GHz. The false colour bar is expressed in Jy. The non uniformity of the noise reflects the non uniform sky scanning strategy of the telescope.}  \label{fig:PSM_nQ}
\end{figure*}

In order to assess the performance of our Bayesian techniques 
under realistic conditions, we use realistic simulations of the QUIJOTE
 experiment Wide Survey \citep{mfiwidesurvey}. The QUIJOTE Wide Survey is observing approximately half the sky at 11, 13, 17 and 19 GHz. These simulations 
have been produced  thanks to the EU RADIOFOREGROUNDS project\footnote{The
RADIOFOREGROUNDS project
aims to combine two unique datasets, the nine \textit{Planck} all-sky (30-857 GHz) maps and the four QUIJOTE Northern sky (10-20 GHz) maps, to provide the best possible characterization of the physical properties of polarized emissions in the microwave domain, together with an unprecedentedly thorough description of the intensity signal. This legacy information will be essential for future sub-orbital or satellite experiments.
See more information in \url{http://www.radioforegrounds.eu/}}, but are not yet public. The simulations make use of 
the
Planck Sky Model \citep[PSM,][]{delabrouille2012},
a global representation of the multi-component sky at frequencies ranging from 1 GHz to 1 THz that
summarizes in a synthetic way as much as possible of our present knowledge of the GHz sky. The PSM is a public code\footnote{http://www.apc.univ-paris7.fr/$\sim$delabrou/PSM/psm.html}  developed mostly by members of the \textit{Planck} collaboration as a simulation tool for \textit{Planck} activities and it makes possible to simulate the sky in total intensity and the $Q$, $U$ Stokes parameters for any experimental configuration in the GHz range. For this work, we choose to simulate the QUIJOTE wide Survey at 11 GHz. Figures~\ref{fig:PSM_Q} and~\ref{fig:PSM_nQ} show the full $Q$ Stokes Wide Survey simulated at 11 GHz and the simulated QUIJOTE instrumental noise for the same Stokes parameter, area and frequency. Table~\ref{tb:params} indicates the main parameters used for this simulation.
 
 \begin{table}
 \centering
\begin{tabular}{lc}
\hline 
\hline 
Frequency (GHz) & 11 \\ 
FWHM (degrees) & 0.85 \\ 
Nside parameter & 246 \\ 
Pixel resolution (arcmin) & 13.74 \\ 
White noise rms (Jy) & 0.386 \\
$\langle \Pi \rangle$ & 0.02 \\
$\sigma_{\Pi}$ & 1.0 \\
\hline 
\hline 
&  \\
\end{tabular} 
\caption{Simulation parameters used for this work.} \label{tb:params}
\end{table}

Formula (\ref{eq:estimatorb}) could be applied to the whole sky, but since the statistical properties of the foregrounds vary strongly with Galactic latitude we prefer to apply the Bayesian estimator locally.
In order to test the method, we compute (\ref{eq:estimatorb}) on flat sky patches, projecting the HEALPix\footnote{Hierarchical Equal Area isoLatitude Pixelation of a sphere,  \url{http://healpix.sf.net}.} simulations described above on  $64\times64$ pixel (that is, a $14.658 \times 14.658$ square degrees area) planar images. 

\begin{figure*}[h]
\subfloat[Stokes Q]{\includegraphics[width=0.5\textwidth]{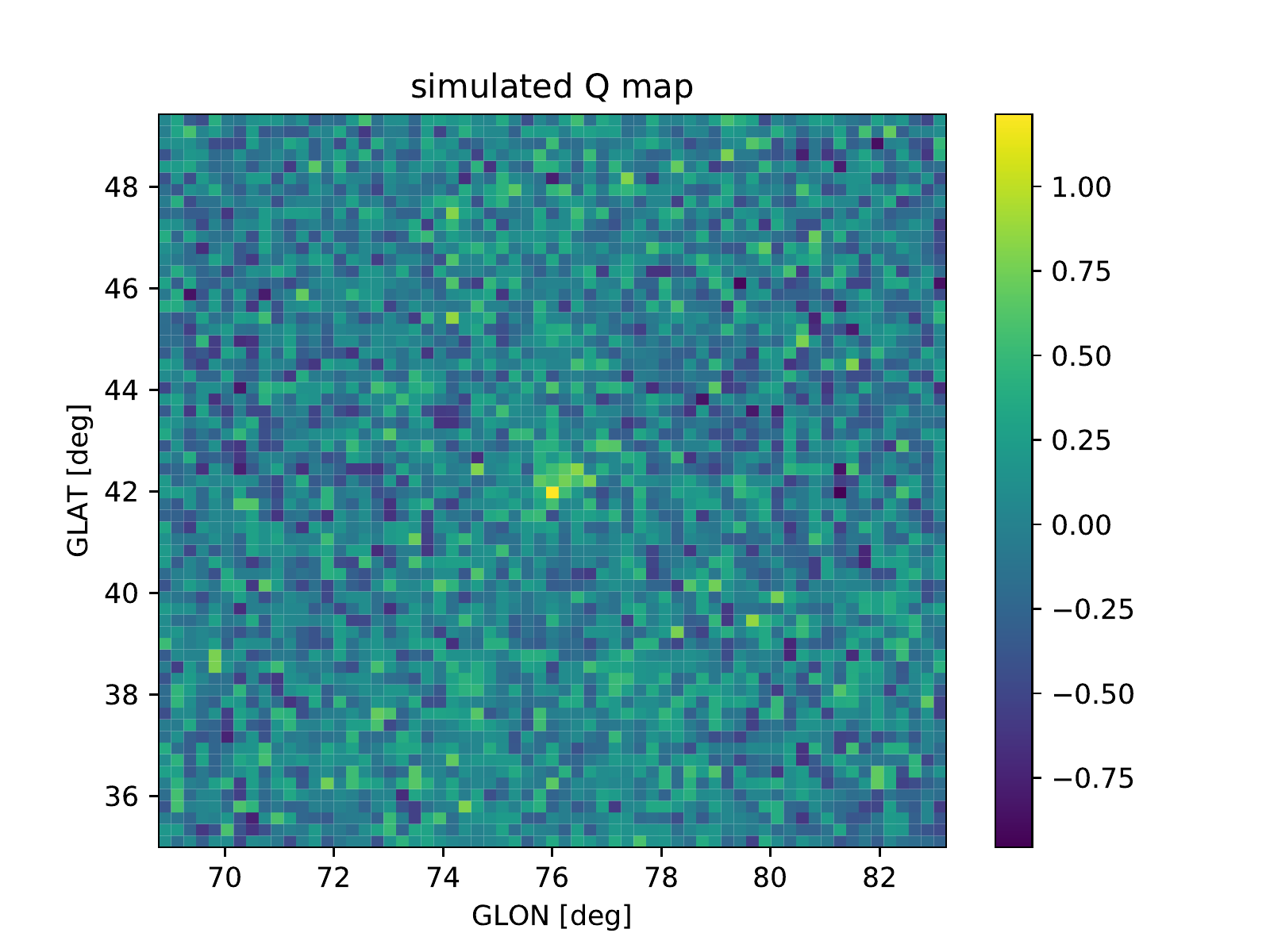}} \qquad
\subfloat[Stokes U]{\includegraphics[width=0.5\textwidth]{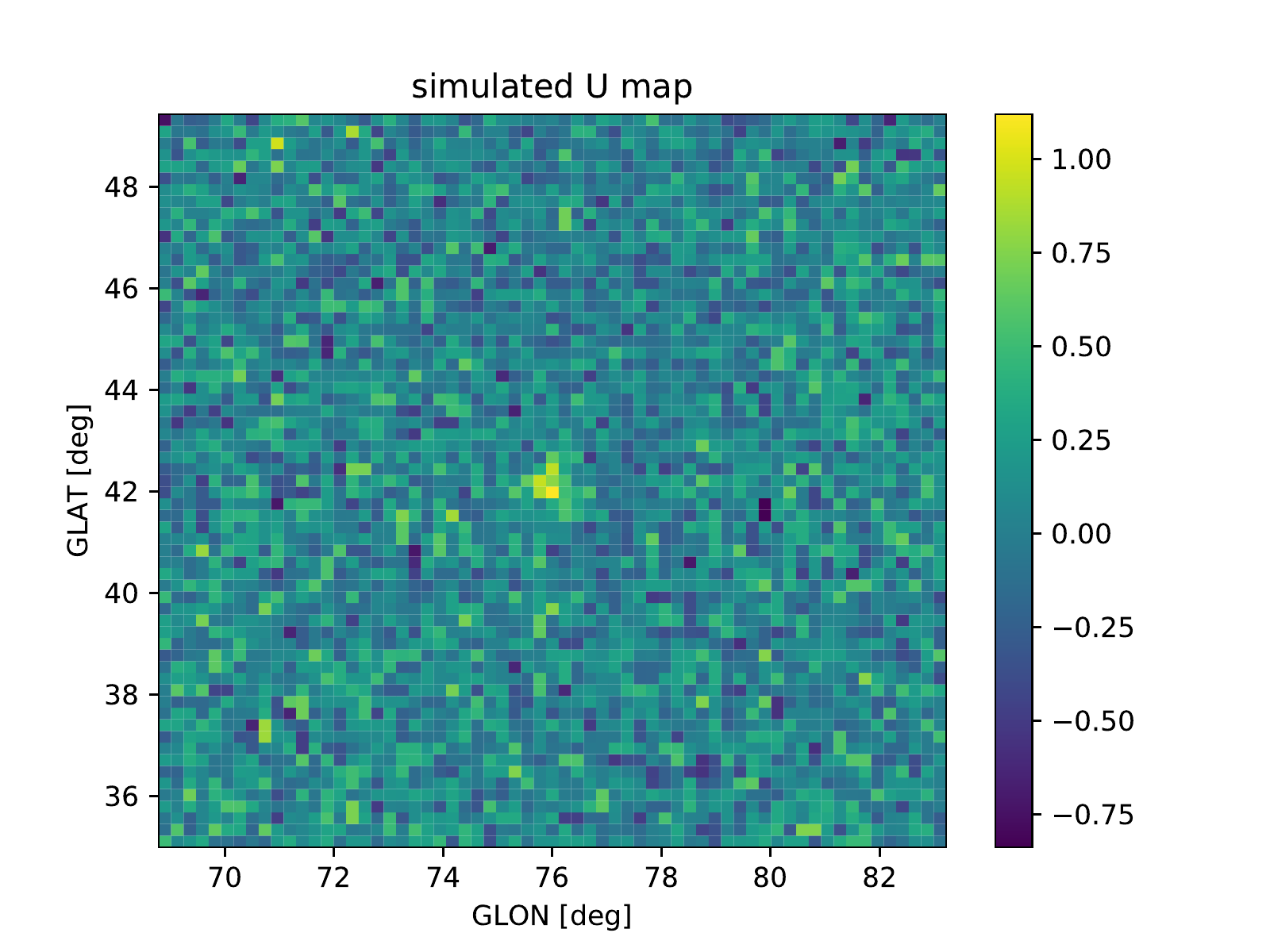}}
\caption{Projected sky patches for one of our simulations. Map units are expressed in Jy.} \label{fig:patches}
\end{figure*}

We have run the estimator on 2000 flat patches as described above. In order to study the effect of the level Galactic contamination, we divide the sky in two areas: 10000 simulations within a `Galactic' band with Galactic latitude $|b|\leq 10^{\circ}$ and 10000 within an external region with $|b| > 10^{\circ}$. The center sky coordinate of each patch is chosen randomly, according to these latitude intervals and inside the simulated Wide Survey observed area (see Figures~\ref{fig:PSM_Q} and~\ref{fig:PSM_nQ}). For each patch, we inject at the center a point source with  the FWHM listed in Table~\ref{tb:params}, a given flux density $S_0$ and polarization fraction $\Pi$ randomly drawn from the log-normal distribution (\ref{eq:lognormal}) with the mean and standard deviation values $\langle \Pi \rangle$  and  $\sigma_{\Pi}$  described in Table~\ref{tb:params}. Please note that the PSM simulations already contain resolved and unresolved polarized point sources apart from the synthetic test sources we are injecting at the central position of each simulated patch. Figure~\ref{fig:patches} shows the $Q$ and $U$ Stokes parameters for one of our simulations. We simulate intensities $S_0$ in ten logarithmically spaced values between 0.1 and 100 Jy (that is, 200 sources --100 of them in the band, 100 of them outside it-- with $S_0=0.1$ Jy, 200 with $S_0=0.2154$ Jy, and so on). In this way we get a sample of sources from moderately faint to extra bright  (and, since the polarization fraction 
follows distribution (\ref{eq:lognormal}) with average $\langle \Pi \rangle = 0.02$, values of $P$ from below 1 mJy to a few tens of Janskys).

\section{Results} \label{sec:results}

\subsection{White noise}  \label{sec:white}

In all the cases analyzed with the white noise simulations, the performance of the Bayesian FF at estimating the source polarization is better than that of the Bayesian Rice method. This was also the case for their non-Bayesian counterparts (Argüeso et al. 2009, etc.). Due to this and taking also into account that the generalization of the Bayesian Rice method to full-sky simulations is far from trivial\footnote{For spatially correlated noise, such as the polarization produced by Galactic foregrounds, the distribution of $P$ is not Ricean any more.}, from now on, we will only compare the Bayesian FF and the FF techniques.

\begin{figure*}[h]
\subfloat[]{\includegraphics[width=0.5\textwidth]{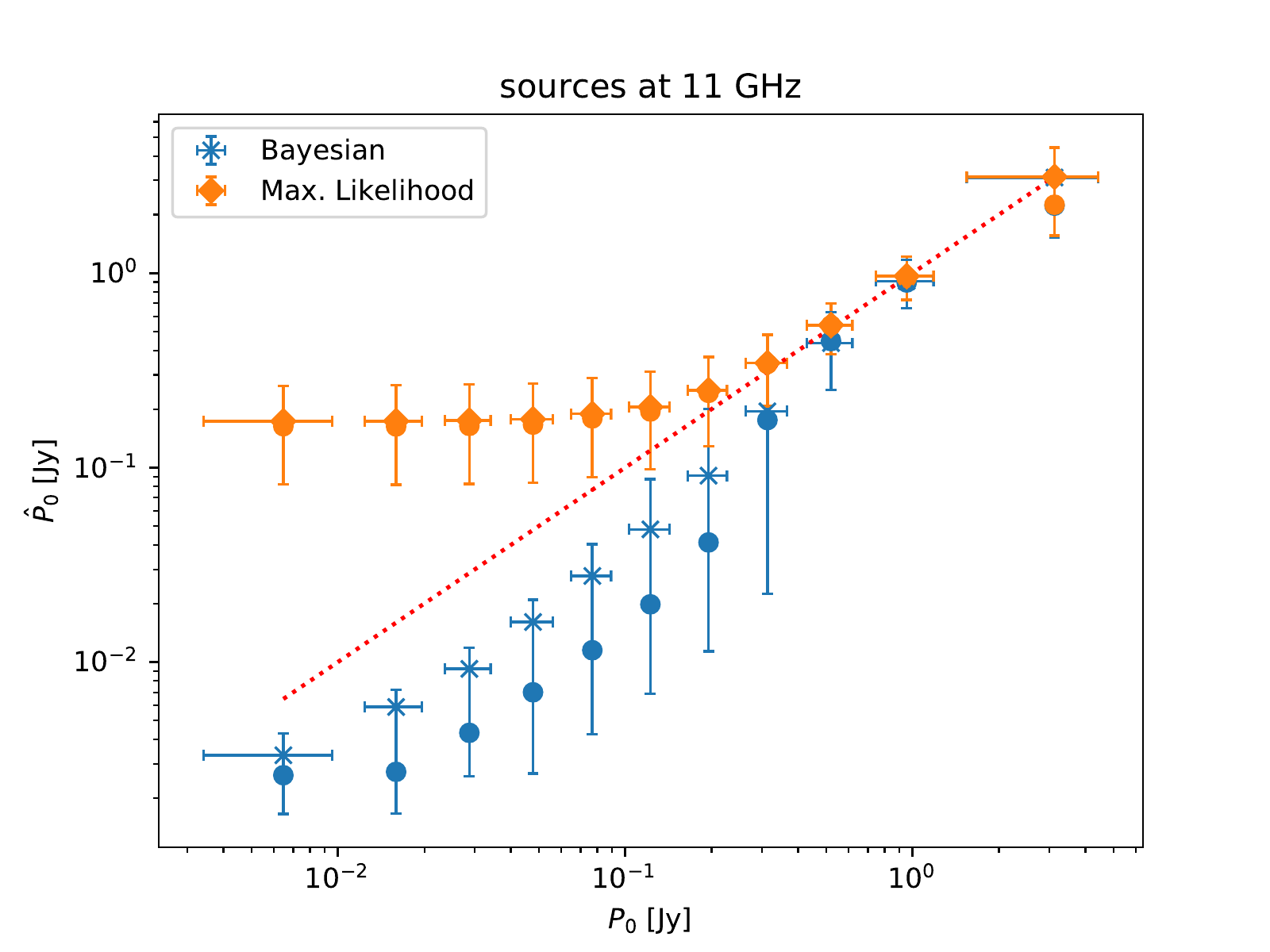}} %\qquad
\subfloat[]{\includegraphics[width=0.5\textwidth]{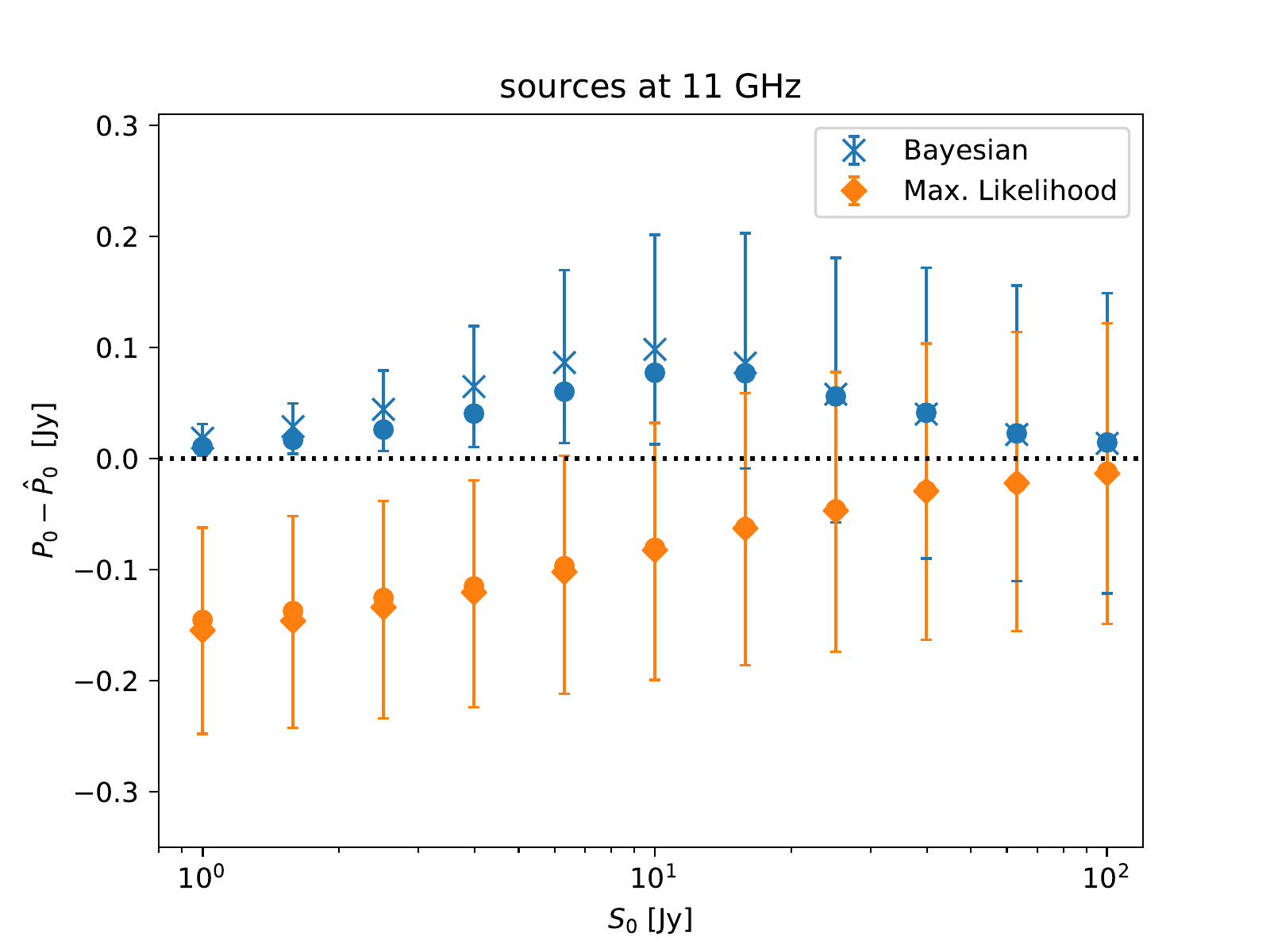}}   
   \caption{\textit{Left}: Binned estimates of the polarized flux density as a function of binned input $P_0$, in Janskys, for the set of 10000 simulated white noise patches.
   \textit{Right}: Error in the estimation $P_0-\hat{P}_0$, as a function of binned input $S_0$ (Jy), for the same set of simulations. Bayesian estimations appear in blue,
   whereas maximum likelihood estimations are shown with orange colour. 
   Error bars show the $68.27\%$ interval of the distribution of results in each case.  Filled circles indicate the median of the distribution; the diamonds (for the MLE) and crosses (for the Bayesian estimator) indicate the average value of the distribution.}  \label{fig:P_white}
\end{figure*}

The left panel of Figure~\ref{fig:P_white}  shows the estimation of the polarized flux density for the 10000 white noise simulations. The results have been binned into eleven logarithmically spaced intervals in input $P_0$. We show in blue the results from the Bayesian estimator (\ref{eq:estimatorb}).
The error bars show the $68.27 \%$ intervals of the corresponding empirical distributions. The filled circles indicate the median value of the distribution of results; the dots indicate the average value of the distribution.
 For comparison, the results of a MLE, which correspond to the third and fourth terms of (\ref{eq:estimatorb}), are shown in orange.
The red dotted line shows the $P=P_0$ line. 
 The MLE is equivalent to the FF technique \citep{argueso09,lopezcaniego09,review}.  Both the Bayesian FF and the MLE estimator work well for highly polarized sources ($P_0 \gtrsim 0.5$ Jy). For input polarization levels below $\sim 0.2$ Jy (which is approximately the rms of the filtered noise of the simulations), however, the MLE reaches a plateau: it is naturally limited by the level filtered noise. The Bayesian estimator, on the contrary, uses the \textit{a priori} information on the $\Pi$ distribution and the knowledge of the source flux density $S$ to predict lower $P$ values. As a matter of fact, the Bayesian FF tends to overcompensate and predict, for the lower end of values of the input $P_0$, polarized fluxes $P_{est} \sim 0.2 P_0$. To see why this happens, we will carry out a short theoretical calculation based on (\ref{eq:estimator})
 . We write the first part of (\ref{eq:estimator}) as a function of $P_0$.
 \begin{eqnarray}
& - & \log{f(Q_0,U_0/Q_i,U_i)}=(\log P_0-\log(\Pi_{med}S_0))^2/2\sigma_{\Pi}^2+ \nonumber \\ 
&& 2 \,\log{(P_0)} +\Sigma_i \displaystyle\frac{(Q_i-Q_0\tau_i)^2}{ 2\sigma_i^2}+\Sigma_i\displaystyle\frac{(U_i-U_0\tau_i)^2}{ 2\sigma_i^2} + K  \nonumber \\
\end{eqnarray} 
\noindent
If we define

\begin{equation}
\Sigma_i \displaystyle\frac{Q_i \tau_i}{\sigma_i^2}=P_1 \cos{\phi_1} \,,
Q_0=P_0 \cos{\phi}
\end{equation} 
and
\begin{equation}
\Sigma_i \displaystyle\frac{U_i \tau_i}{\sigma_i^2}=P_1 \sin{\phi_1} \,,
U_0=P_0 \sin{\phi}
\end{equation} 
 
\noindent the previous formula can be expressed in terms of $P_0$, $P_1$ and the polarization angles

 \begin{eqnarray}
% \begin{aligned}
 &-& \log{f(Q_0,U_0/Q_i,U_i)}=(\log P_0-\log(\Pi_{med}S_0))^2/2\sigma_{\Pi}^2+ \nonumber \\ 
 && 2 \,\log{(P_0)} 
 -P_0 P_1 \cos{(\phi-\phi_1)}+  P_0^2\, \Sigma_i\displaystyle\frac{\tau_i^2}{ 2\sigma_i^2} + K_2
 \nonumber \\
 \end{eqnarray}  
 \noindent
 with
 \begin{equation} 
 \Sigma_i\, \displaystyle\frac{Q_i^2+U_i^2}{ 2\sigma_i^2}+K=K_2
 \end{equation} 
 Taking the partial derivatives of (18)  with respect to $P_0$ and $\phi$ and equating them to zero, we obtain the estimators $\hat{P_0}$ and $\hat{\phi}$ that minimize minus the log-posterior. It can be easily seen that $\hat{\phi}=\phi_1$ and $\hat{P_0}$ is the solution of the following equation
 \begin{equation}
 (\log \hat{P_0}-\log(\Pi_{med}S_0))/\sigma_{\Pi}^2+2-\hat{P_0}\,P_1+\hat{P_0}^2\, \Sigma_i\, \displaystyle\frac{\tau_i^2}{\sigma_i^2}=0
 \end{equation} 
 \noindent
 This equation is very interesting: if we assume that $ \hat{P_0}\,P_1 \ll 1$ and $\hat{P_0}^2\, \Sigma_i\, \displaystyle\frac{\tau_i^2}{\sigma_i^2} \ll 1$, that is, the source polarization is much lower than the noise, the estimation will be dominated by the Bayesian prior and neglecting the last two terms in (20), we find a constant value for the estimator, independently of the data,

 \begin{equation}
 \hat{P_0}=\Pi_{med}\,S_0\, e^{-2\sigma_{\pi}^2}
 \end{equation}
 \noindent
For a lognormal distribution
\begin{equation}
\Pi_{med} = \langle \Pi \rangle   \exp\left(-\sigma_{\Pi}^2/2\right),
\end{equation}
\noindent
Taking into account the values of $\langle \Pi \rangle$ and $\sigma_{\Pi}$ given in Table~\ref{tb:params}, $\Pi_{med}=0.012$ and
we obtain 
\begin{equation} 
 \hat{P_0}=0.00164\, S_0.
 \end{equation}
 \noindent In order to check these theoretical results, we have carried out $10000$ simulations with $S_0=1 $ Jy. In this case, the noise, $ \sigma=0.386$ Jy, is much higher than the source polarization. We find, for all our simulations, the estimated value $\langle \hat{P}_0 \rangle = 0.00166 \pm 0.00002$, compatible with the calculation above. 
 
 Though the estimator is constant in this case, this value is closer to the real value than that obtained by using the matched filter, which is completely dominated by the noise.
 
 For higher values of $S_0$, e. g \, $S_0=10$  Jy, there are around $5000$ simulations, corresponding to the lower polarizations,  that produce an estimator close to the default value $0.016-0.020$ Jy. For higher values of the real polarization, there is a combination of the prior and the matched filter terms in the solution of (20). At any rate, the performance of the Bayesian FF is better than that of the plain FF.

 \bigskip
 
  The right panel of 
 Figure~\ref{fig:P_white} shows the polarized flux estimation error\footnote{Defined as $P_0-\hat{P}_0$,
 where $P_0$ is the input value and $\hat{P}_0$ the estimated value of the polarization of the source, either through the Bayesian method or through the MLE.}   as a function of the input flux density (temperature) of the source $S_0$. For low flux densities, the figure  shows both the systematic overestimation of the MLE, due to the noise limit, and the underestimation of the Bayesian FF estimator, due to the reasons discussed above. In absolute terms the statistical error of the MLE is much larger than that of the Bayesian FF estimator in the low flux density regime. There is an interesting interval at intermediate flux densities ($\sim 10$ Jy) at which the errors of the MLE and  Bayesian FF estimator are of the same order, but in opposite directions. The Bayesian estimator seems to reach a plateau (i.e., is noise-limited) around  $P_0 \sim 10$ mJy, a order of magnitude below in polarized flux than the MLE.   

\begin{figure}[h]
  \centering
   \includegraphics[width=\columnwidth]{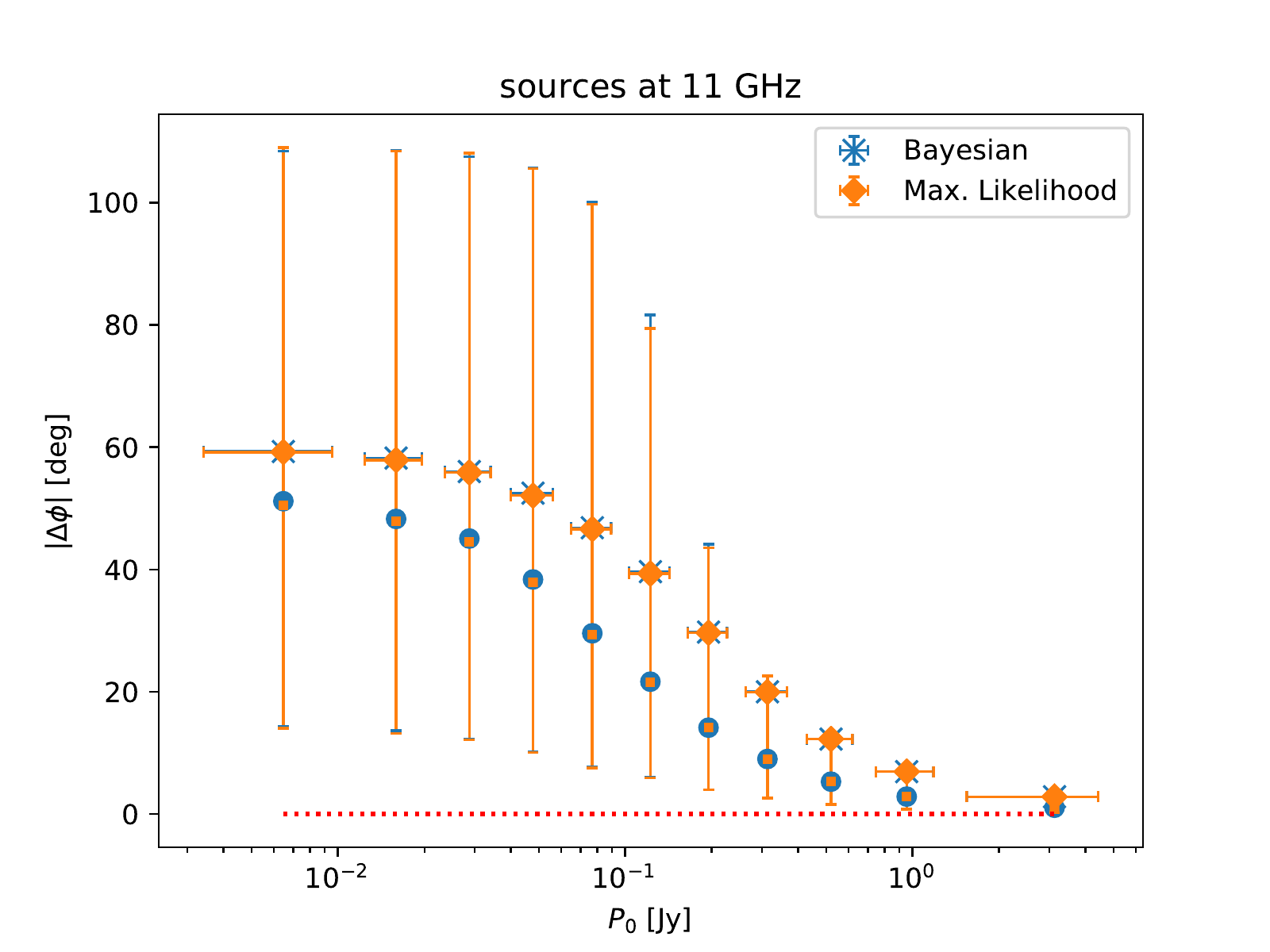}
   \caption{Binned estimates of $|\Delta \phi| = |\phi_0-\hat{\phi}_0|$ as a function of binned input $P_0$, in Janskys, for the set of 10000 simulated white noise images. Bayesian estimations are marked with blue crosses and error bars,
   whereas maximum likelihood estimations are shown with orange diamonds and error bars. The median values are indicated by means of large blue filled circles (Bayesian estimator) and small organge squares (MLE).}   \label{fig:phi_white}
\end{figure}

Figure~\ref{fig:phi_white}
shows the absolute polarization angle error,
\begin{equation}
|\Delta \phi| = |\phi_0 - \hat{\phi}_0|,
\end{equation}
\noindent
where $\phi_0$ and $\hat{\phi}_0$ are the input and estimated polarization angles (in degrees), as a function of the input polarization of the source $P_0$. The figure shows that there is little difference between the Galactic and extragalactic areas, and between the Bayesian estimator and the MLE. This is not a surprise, since the priors in (\ref{eq:estimatorb}) are constant with respect to $\phi_0$, that is, the Bayesian estimator and the MLE should perform similarly, as it is the case.

\subsection{Full sky simulations}  \label{sec:color}

\begin{figure*}[h]
\subfloat[]{\includegraphics[width=0.5\textwidth]{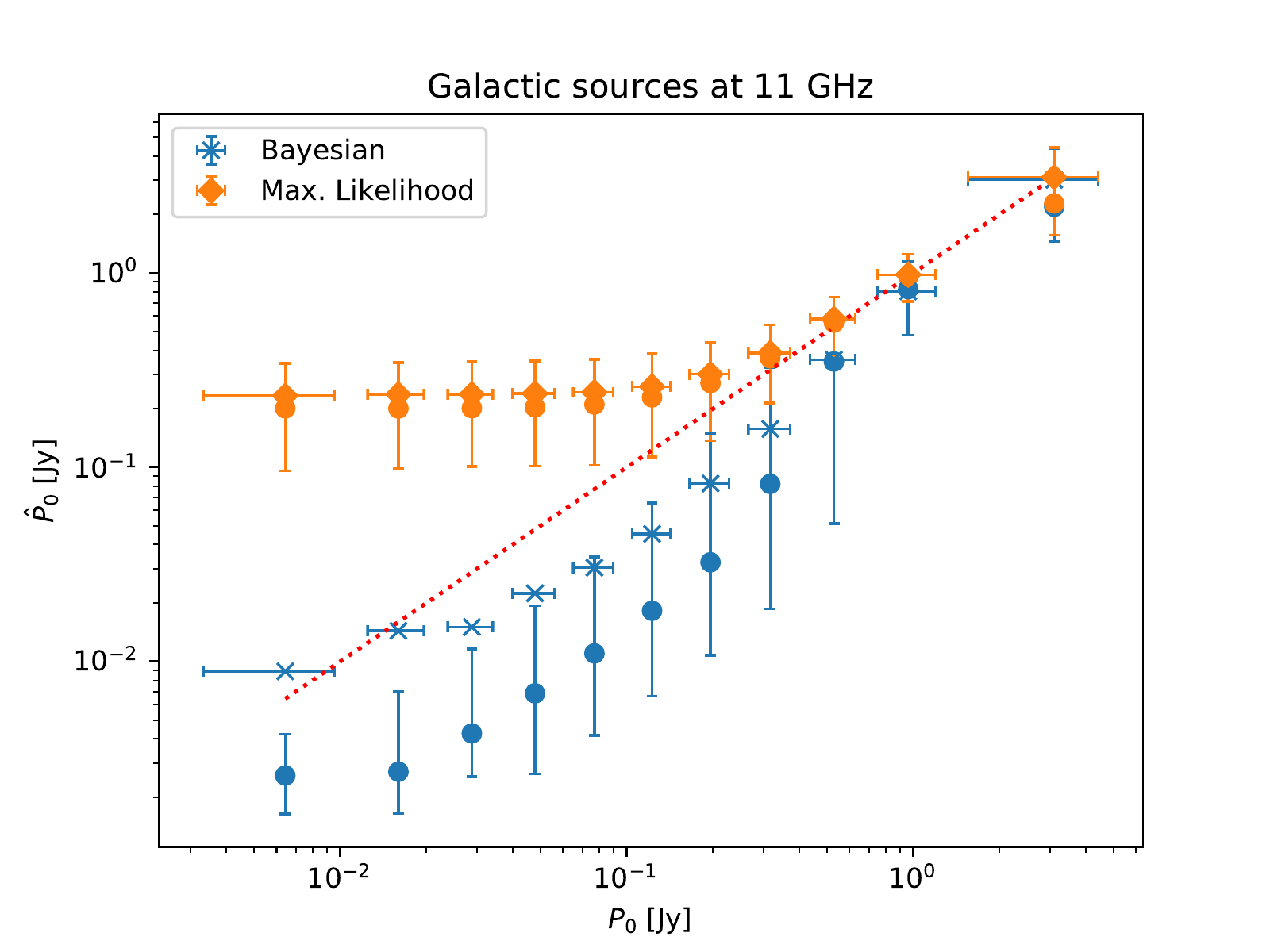}} \qquad
\subfloat[]{\includegraphics[width=0.5\textwidth]{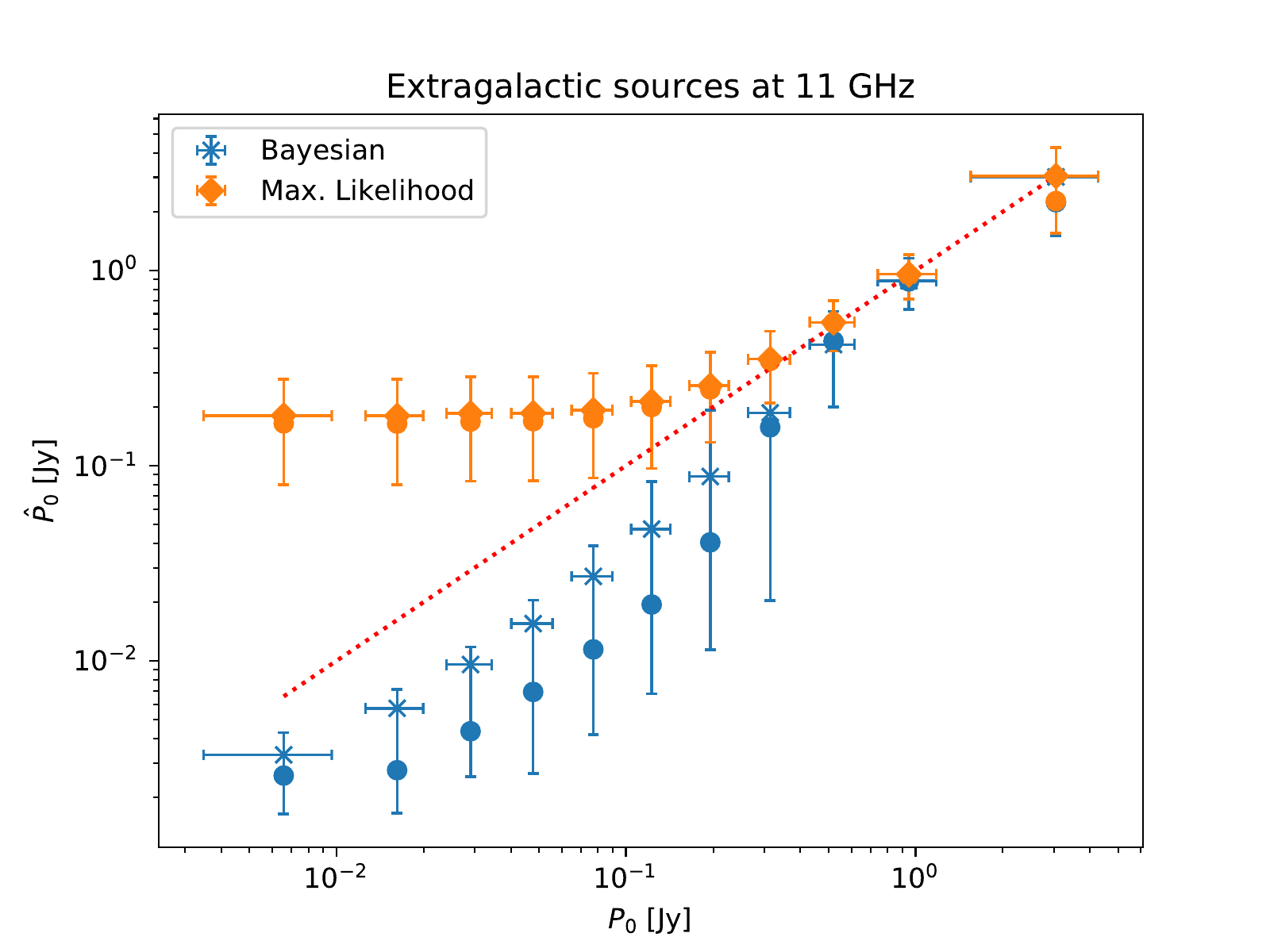}}
\caption{Binned estimates of the polarized flux density as a function of binned input $P_0$, in Janskys, for the set of 10000 simulated QUIJOTE sky patches  with $|b| \leq 10^{	\circ}$ (left) and the
set of 10000 simulated QUIJOTE sky patches  with $|b| > 10^{	\circ}$ (right)
 at 11 GHz. Bayesian estimations appear in blue,
   whereas maximum likelihood estimations are shown with orange colour. Error bars show the $68.27\%$ interval of the distribution of results in each case. Filled circles indicate the median of the distribution; the diamonds (for the MLE) and crosses (for the Bayesian estimator) indicate the average value of the distribution.} \label{fig:Pest}
\end{figure*}

Figure~\ref{fig:Pest}  shows the estimation of the polarized flux density
 for a) the 10000 `Galactic' ($|b|\leq 10^{\circ}$) and
b) the 10000 `extragalactic' ($|b|>10^{\circ}$) simulated QUIJOTE sky patches. The results have been binned into eleven logarithmically spaced intervals in input $P_0$. We show in blue the results from the Bayesian estimator (\ref{eq:estimatorb}).
The error bars show the $68.27 \%$ intervals of the corresponding empirical distributions. 
 For comparison, the results of a MLE, which correspond to the third and fourths terms of (\ref{eq:estimatorb}), are shown in orange. The MLE is equivalent to the FF technique \citep{argueso09,lopezcaniego09,review}.  As it happened in the case of the white noise simulations (Section~\ref{sec:white}), for low flux density sources the MLE estimator
 reaches a plateau dominated by the noise level (higher for Galactic than for extragalactic sources). The Bayesian estimator reaches a similar plateau at much lower polarized fluxes, again in the $\sim 10$ mJy regime instead of the $\sim 100$ mJy regime of the MLE. Please note, however, that the distribution of the estimated $\hat{P}_0$ by means of the Bayesian estimator becomes more and more skewed as $P_0$ decreases\footnote{This can be quickly seen by the growing differences between the median and the average values of the distribution, as shown in the Figure.}.

\begin{figure*}[h]
\subfloat[]{\includegraphics[width=0.5\textwidth]{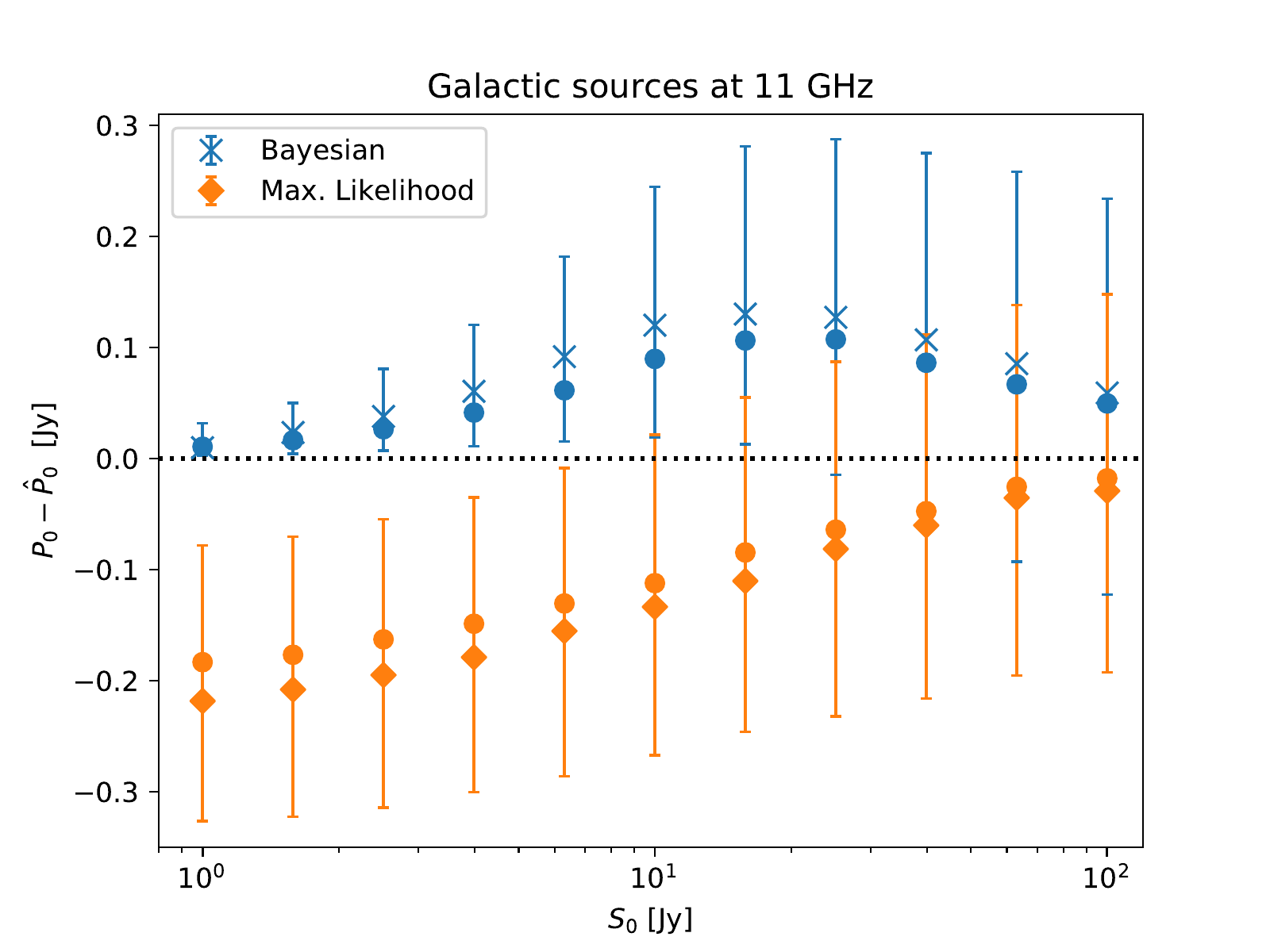}} \qquad
\subfloat[]{\includegraphics[width=0.5\textwidth]{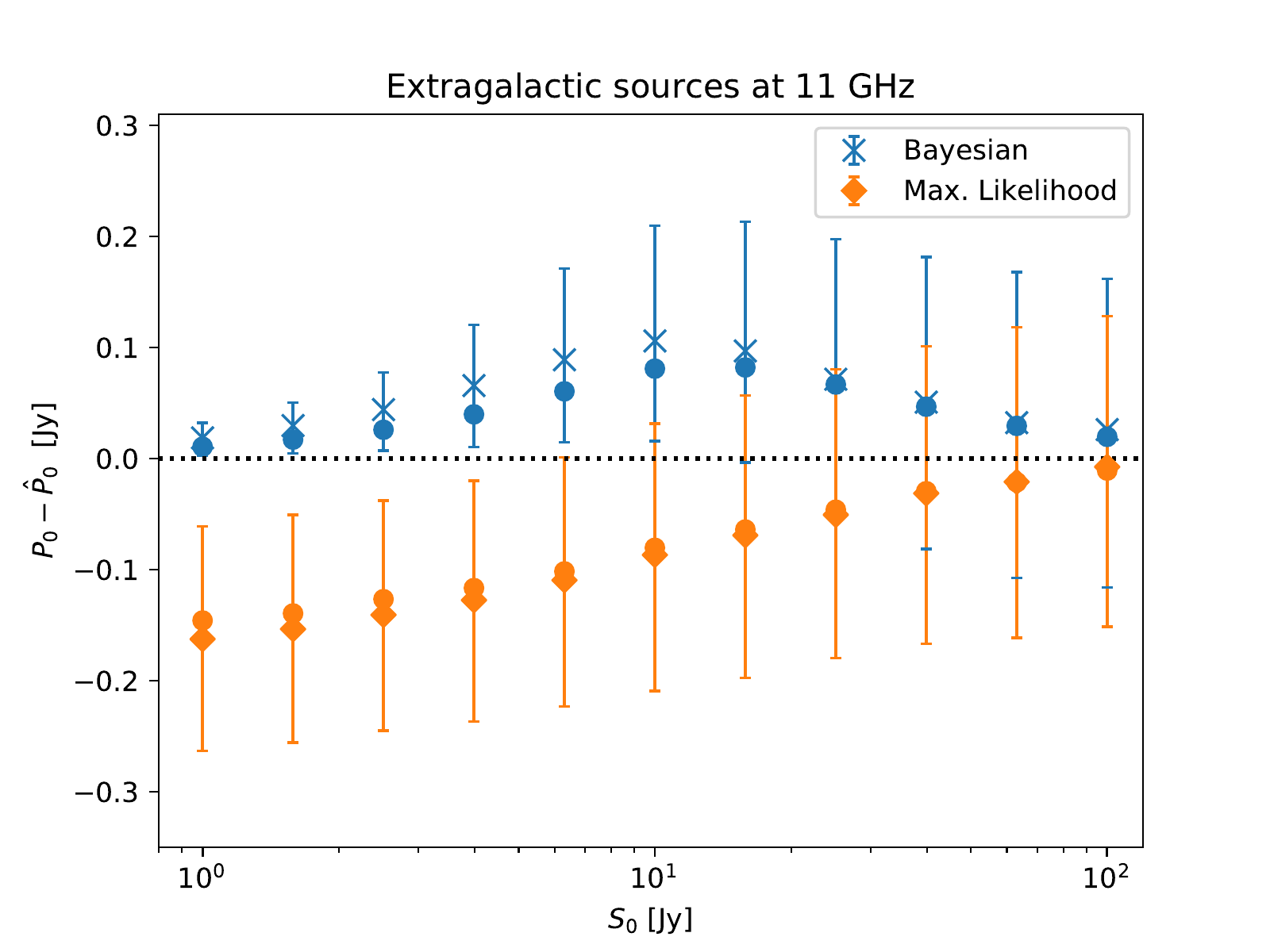}}
\caption{Error in the estimation $P_0-\hat{P}_0$ (Jy), as a function of binned input $S_0$ (Jy), for the set of 10000 simulated QUIJOTE sky patches  with $|b| \leq 10^{	\circ}$ (left) and the
set of 10000 simulated QUIJOTE sky patches  with $|b| > 10^{	\circ}$ (right)
 at 11 GHz. Bayesian estimations appear in blue,
   whereas maximum likelihood estimations are shown with orange colour. 
   Error bars show the $68.27\%$ interval of the distribution of results in each case. Filled circles indicate the median of the distribution; the diamonds (for the MLE) and crosses (for the Bayesian estimator) indicate the average value of the distribution.}
    \label{fig:Perr}
\end{figure*}

\begin{figure*}[h]
  \centering
   \includegraphics[width=\textwidth]{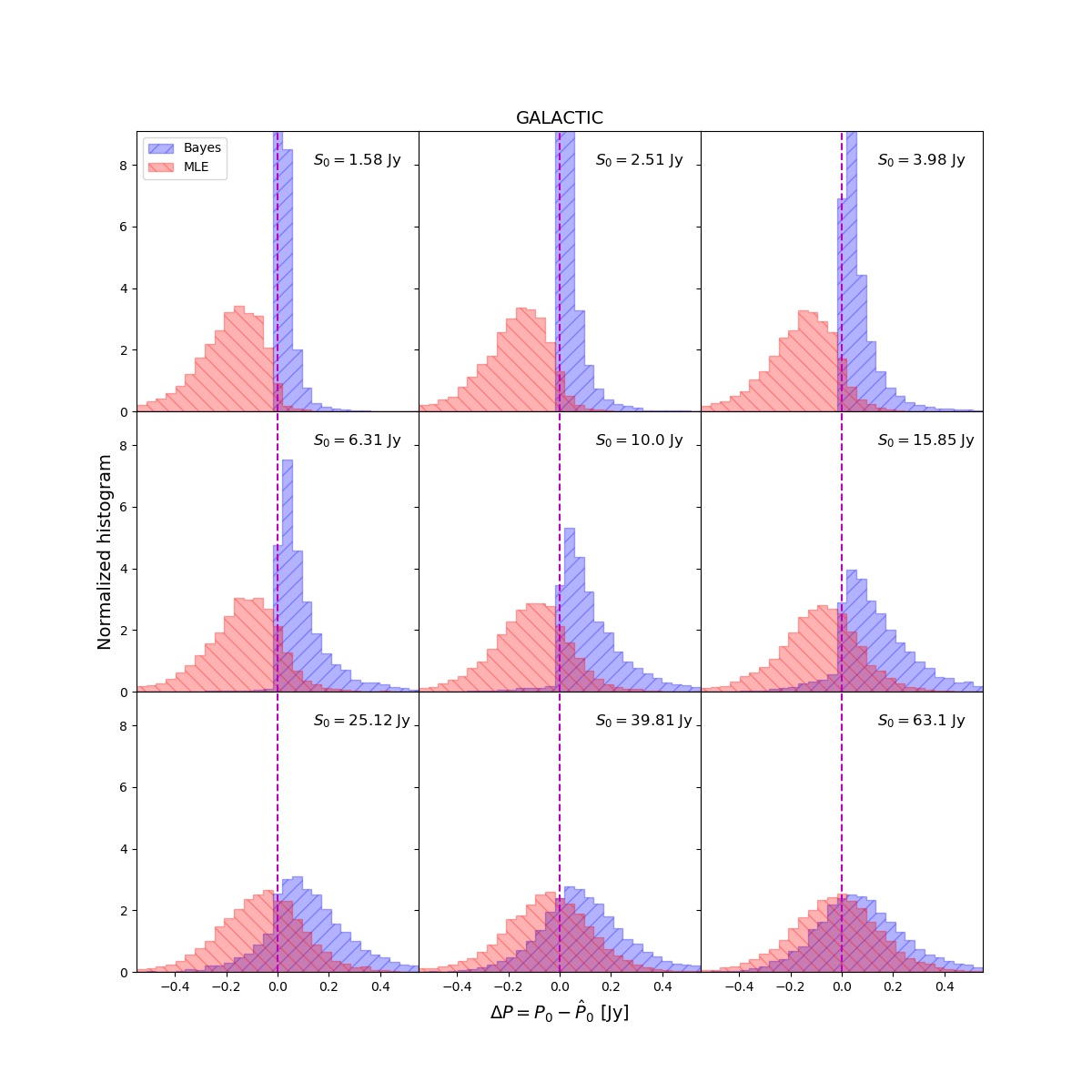}
   \caption{Normalized histogram of the difference $\Delta P$ between the input polarization $P_0$ and the estimated polarization for the Bayesian estimator (blue, /) and the MLE (red, \textbackslash), for sources located within the Galactic band $|b| \leq 10^{\circ}$, and for nine different values of the input total flux density $S_0$.}  \label{fig:Pdiff_gal}
\end{figure*}

\begin{figure*}[h]
  \centering
   \includegraphics[width=\textwidth]{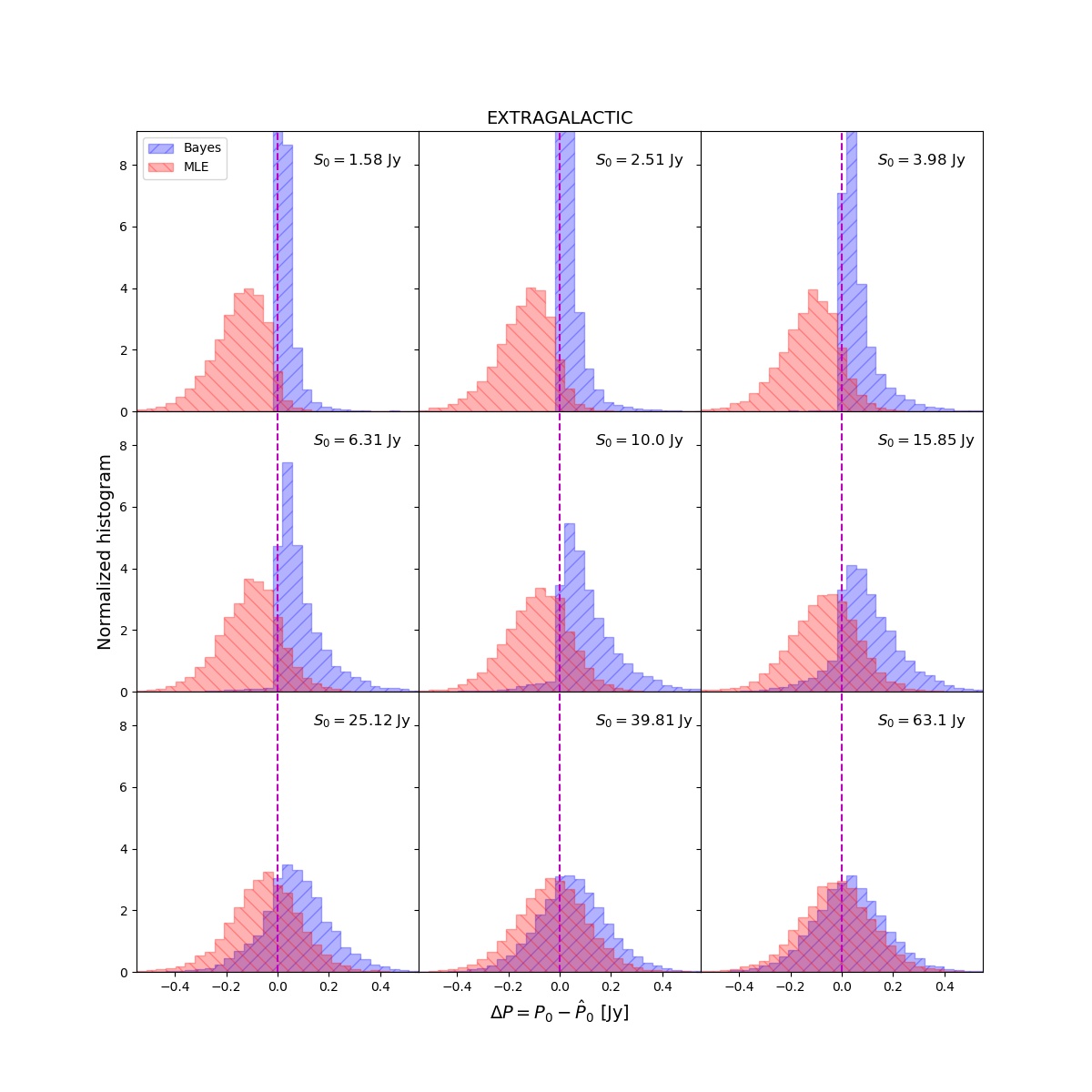}
   \caption{Normalized histogram of the difference $\Delta P$ between the input polarization $P_0$ and the estimated polarization for the Bayesian estimator (blue, /) and the MLE (red, \textbackslash), for sources located outside the Galactic band $|b| > 10^{\circ}$, and for nine different values of the input total flux density $S_0$.}  \label{fig:Pdiff_extra}
\end{figure*}

 Figure~\ref{fig:Perr} shows the error of the estimation of $P$ as a function of the input flux density of the sources (in Janskys). 
As in the case of white noise, the MLE estimator tends to overestimate the polarized flux of faint sources whereas the Bayesian FF tends to underestimate it. This error is a systematic bias that tends to a constant value in relative terms, but  decreases to zero Janskys in absolute terms for $S_0 \rightarrow 0$. Error bars are smaller for extragalactic sources than for Galactic sources, which are embedded in more intense foreground emission.
 Figures~\ref{fig:Pdiff_gal} and~\ref{fig:Pdiff_extra} show the normalized histograms of the difference $\Delta P$ between the input polarization $P_0$ and the estimated polarization $P$,
\begin{equation}
\Delta P = P_0 - \hat{P}_0,
\end{equation}
\noindent
for eleven different values of the total (Stokes I) flux density $S_0$. `Galactic' sources are shown in Figure~\ref{fig:Pdiff_gal}  and `extragalactic' sources are shown in Figure~\ref{fig:Pdiff_extra}. The estimation $\hat{P}_0$ has been obtained with the Bayesian Filtered Fusion method introduced in this paper (in blue) and the MLE (in red color). For bright sources ($S_0 > 10$ Jy) the histograms are approximately symmetric and centered around $\Delta P=0$, but for fainter sources the MLE histograms are skewed to the left, showing the same kind of overestimation already observed in Figure~\ref{fig:Pest}. The histograms for the Bayesian estimator, however, are skewed to the right but much narrower than the MLE histograms, 
which indicates that the Bayesian estimator predicts the  polarization of a source with a smaller margin of error. Both types of error, MLE-overestimation and Bayesian FF-underestimation, must be dealt with in CMB polarization experiments, but the amount of bias is signficantly smaller for the Bayesian FF estimator.

\begin{figure*}[h]
\subfloat[]{\includegraphics[width=0.5\textwidth]{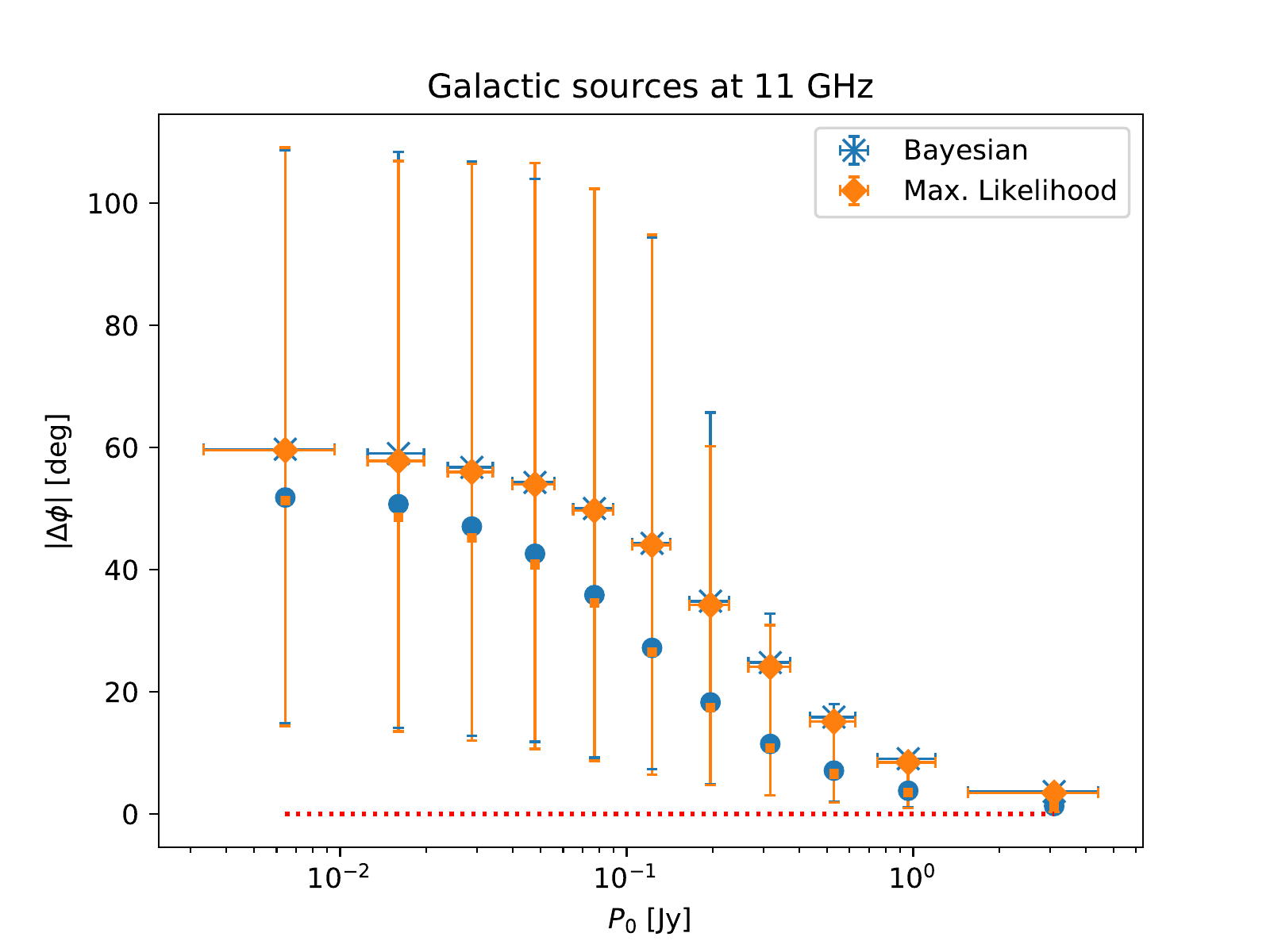}} \qquad
\subfloat[]{\includegraphics[width=0.5\textwidth]{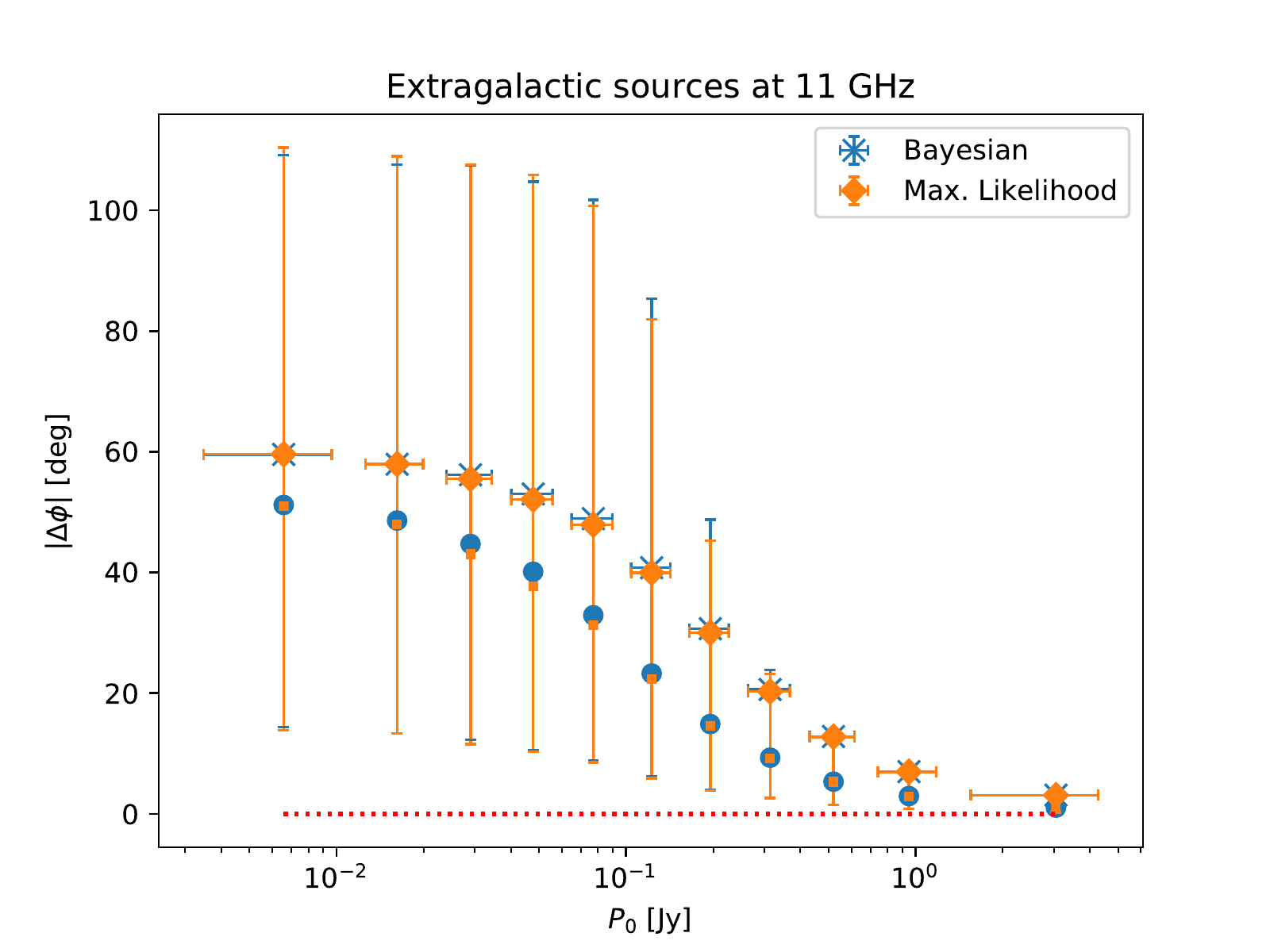}}
\caption{Binned estimates of $|\Delta \phi| = |\phi_0-\hat{\phi}_0|$ as a function of binned input $P$, in Janskys, for the set of 500 simulated QUIJOTE sky patches  with $|b| \leq 10^{	\circ}$ (left) and the
set of 500 simulated QUIJOTE sky patches  with $|b| > 10^{	\circ}$ (right)
 at 11 GHz. Bayesian estimations are marked with blue crosses,
   whereas maximum likelihood estimations are marked with orange diamonds. Median values are indicated with large blue filled circles and small orange filled squares, respectively. } \label{fig:phi}
\end{figure*}

Finally, Figure~\ref{fig:phi} shows the absolute polarization angle error, as a funcion of the input polarization $P_0$.
The figure shows that there is little difference between the Galactic and extragalactic areas, and between the Bayesian estimator and the MLE. This is not a surprise, since the priors in (\ref{eq:estimatorb}) are constant with respect to $\phi_0$, that is, the Bayesian estimator and the MLE should perform similarly, as it is the case.

\subsection{A note on the robustness of the Bayesian estimator}

\begin{figure}[h]
  \centering
   \includegraphics[width=\columnwidth]{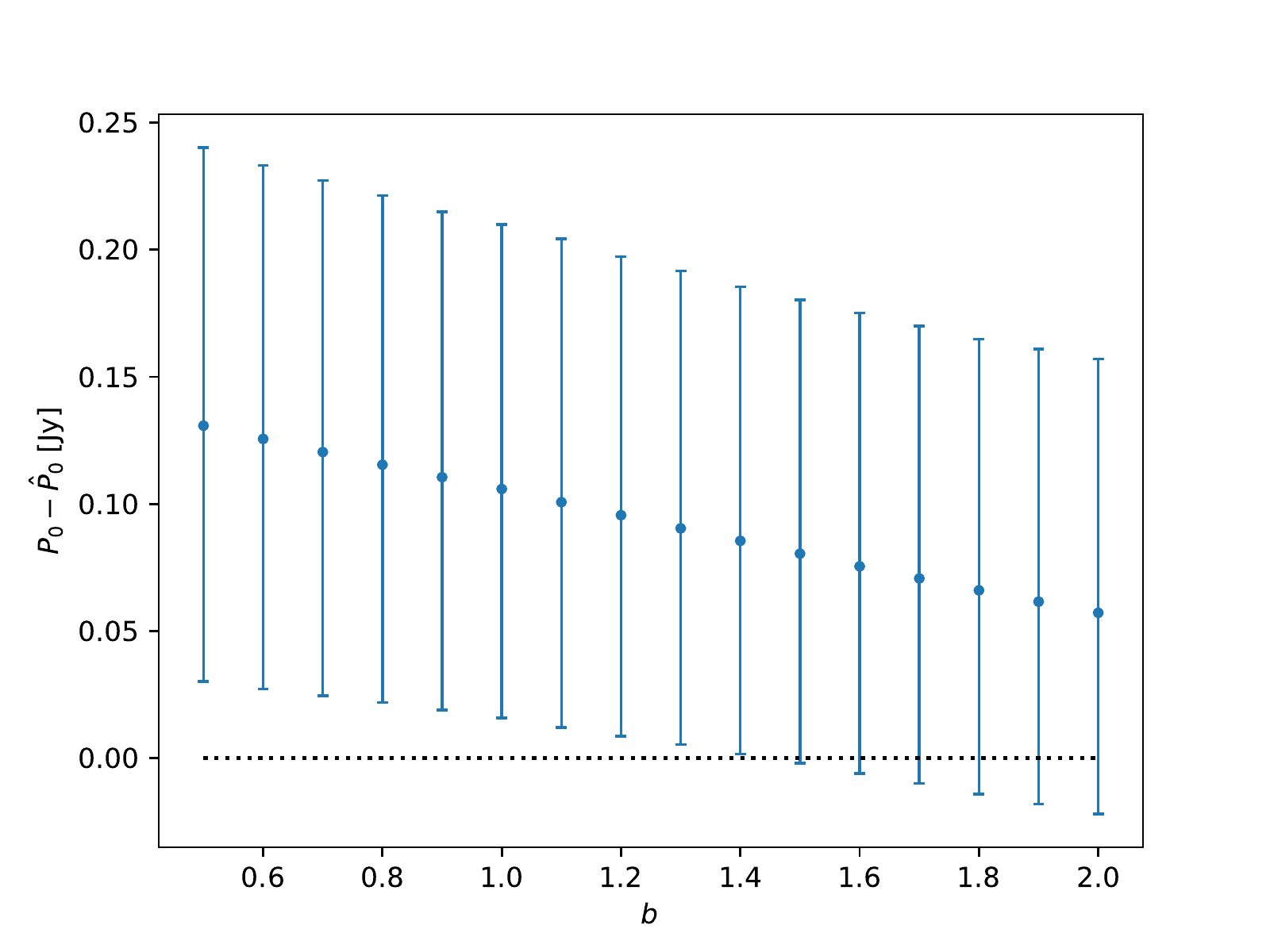}
   \caption{Error in the estimation of the total polarization of a test source with $S_0=10$ Jy as a function of the bias factor $b$ (defined as $\Pi_{med}^b= b \, \Pi_{med}$) affecting the Bayesian prior on 
   $\Pi_{med}$.}  \label{fig:Pdiff_robust}
\end{figure}

Every time some prior information is used in the Bayesian framework the inevitable question arises: what is the effect of a wrong guess of the prior in the estimation? In order to shed some light on this we have re-analyzed the one hundred simulations of `extragalactic' sources with flux density $S_0=10$ Jy\footnote{We have chosen this particular flux density value because according to Figure~\ref{fig:Pdiff_extra} it marks the flux density for which the Bayesian estimator begins to outperform the MLE.}. Instead of using the correct value of the median polarization fraction $\Pi_{med}$ in equation (\ref{eq:estimatorb}) we use a biased parameter $\Pi_{med}^b= b \, \Pi_{med}$ with $b=[0.5,0.6,\ldots,2.0]$, that is, we have tested what happens if our guess of the median polarization fraction is wrong by a factor from $50\%$ to $200\%$. Figure~\ref{fig:Pdiff_robust} shows the average estimation error $P_0-\hat{P}_0$ as a function of the bias factor $b$. Error bars show the $68.27 \%$ intervals of the corresponding empirical distributions.  The figure shows that the average error of the Bayesian estimator varies smoothly with the bias in the prior. For comparison, for the same simulations the MLE produces a (bias independent, as the maximum likelihood estimator does not use prior information) value for the error $P_0-\hat{P}_0 = -0.22 \pm  0.12$, larger than the Bayesian estimation (for this particular value of $S_0$) even when the prior is wrong by a factor of two. 

Another potential source of bias is the uncertainty on the true flux density of the source. The estimators in equations (\ref{eq:estimator}) and (\ref{eq:estimatorb}) 
depend implicitly on an \textit{a priori} knowledge of the source flux density $S_0$ through the factor $\mu_1 = \mu + \log(S_0)$. In the previous tests we have assumed that $S_0$ is known with arbitrary precision, but in practice this will  not be the case. In a real experiment one expects to know some reasonable estimation $\hat{S}_0$ of the true flux density of the source. In a typical CMB experiment setting the difference between $S_0$ and $\hat{S}_0$ will be relatively small (at least in comparison with the relative difference between $P_0$ and $\hat{P}_0$), but not zero. 
The uncertainty on the source flux density can bias 
the estimators (\ref{eq:estimator}) and (\ref{eq:estimatorb}) even if the distribution of $\hat{S}_0$ is symmetric around $S_0$, as $S_0$ enters the estimators in a non-linear fashion. Moreover, one expects the uncertainty in $S$ to increase the statistical error of the estimators. 

In order to test the effect of the uncertainty on $S$ on our Bayesian estimator, we have conducted a new batch of 10000 simulations in the same fashion as described in Section~\ref{sec:color}. The analysis follows the same pipeline as described above, but every time we compute the estimator  (\ref{eq:estimatorb}) we introduce a random photometric error in $S_0$. These photometric errors follow a Gaussian distribution of standard deviation $\sigma = 0.3$ Jy, a little smaller than the QUIJOTE simulation noise rms level\footnote{We assume that the rms of the  photometric errors has been lowered by means of some filtering scheme, such as a matched filter or a Mexican Hat Wavelet, or any other suitable signal processing technique.  Then the $\sim 0.3$ Jy uncertainty becomes a more realistic approximation of error in the determination the flux density of compact sources in the QUIJOTE Wide Survey.}.

\begin{figure*}[h]
\subfloat[]{\includegraphics[width=0.5\textwidth]{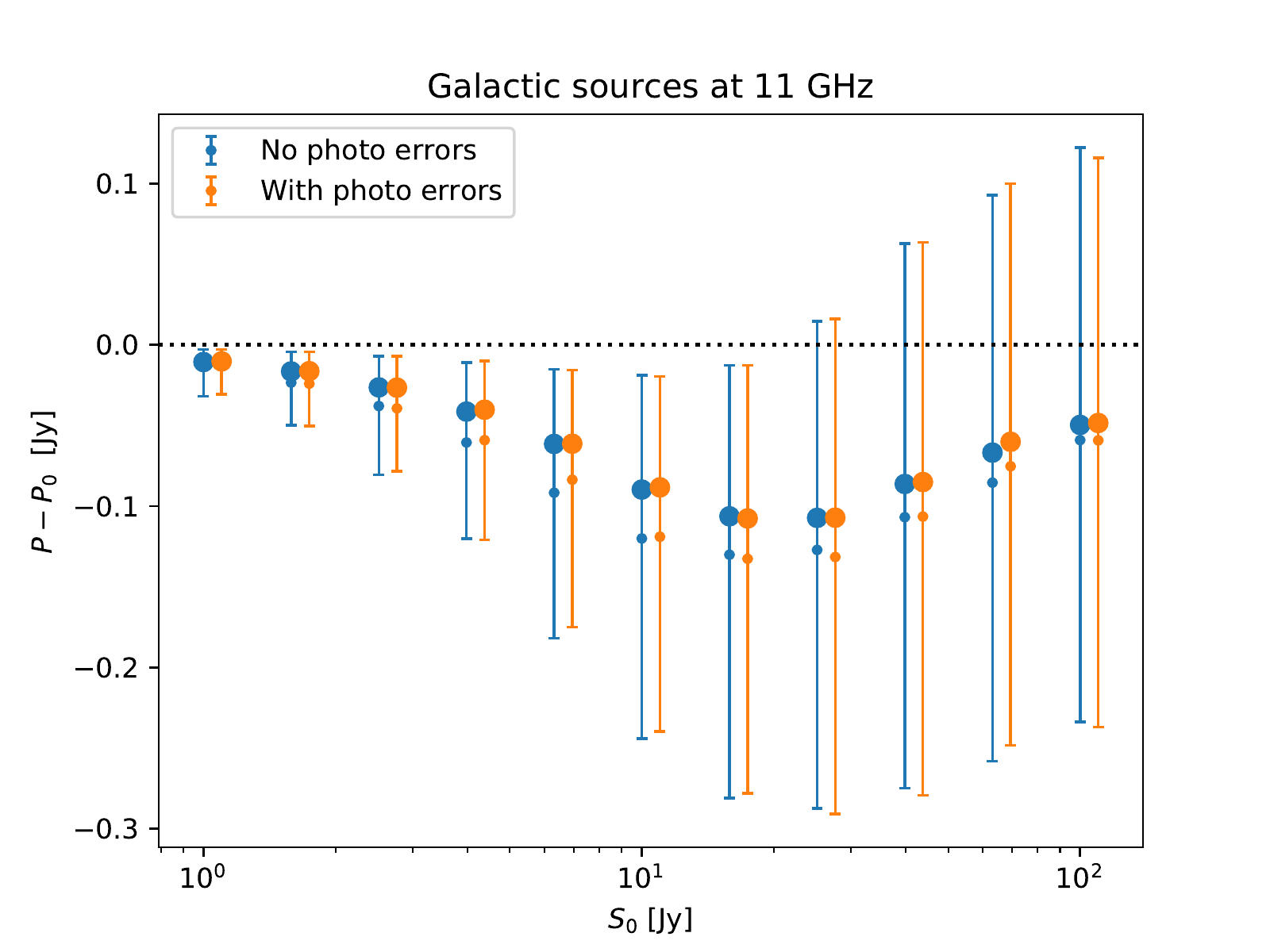}} \qquad
\subfloat[]{\includegraphics[width=0.5\textwidth]{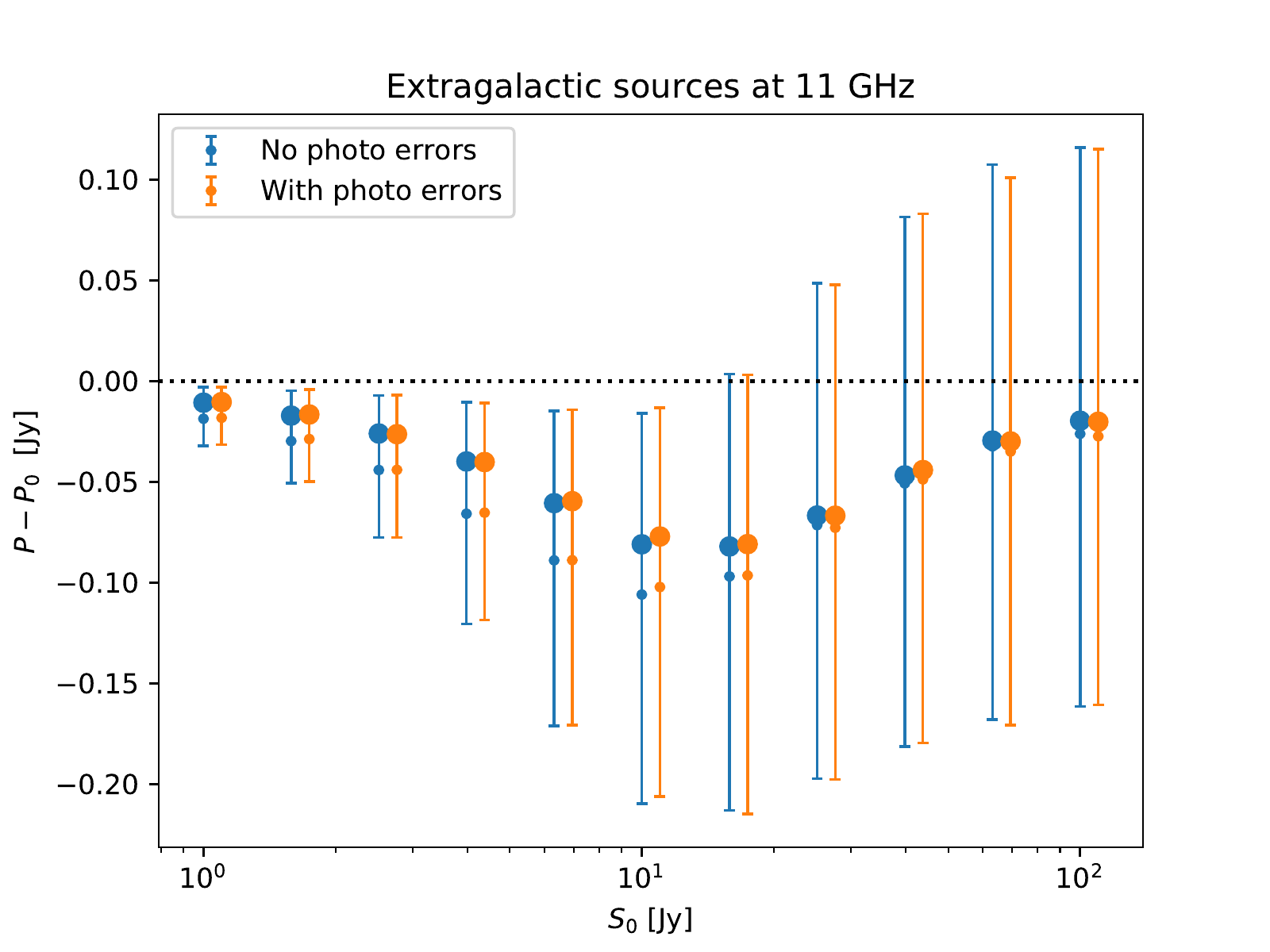}}
\caption{Error in the estimation $P-P_0$ (Jy), as a function of binned input $S_0$ (Jy), for the set of 10000 simulated QUIJOTE sky patches  with $|b| \leq 10^{	\circ}$ (left) and the
set of 10000 simulated QUIJOTE sky patches  with $|b| > 10^{	\circ}$ (right)
 at 11 GHz. Bayesian estimations for a perfect photometry of the source total flux density $S_0$ appear in blue,
   whereas Bayesian estimations including a $0.3$ Jy uncertainty in $S_0$ are shown with orange. 
   Error bars show the $68.27\%$ interval of the distribution of results in each case. Filled circles indicate the median of the distribution; dots indicate the average value of the distribution. Orange points and lines have been slightly displaced to the right in order to make the figure more readable.
   } \label{fig:Perr_photo}
\end{figure*}

Figure~\ref{fig:Perr_photo} shows the average error of the estimation of the polarization of our simulated sources comparing the two cases: if the source flux density $S_0$ is perfectly known in advance (blue dots and error bars) or if a $0.3$ Jy photometric uncertainty is present in the analysis (orange dots and bars, slightly displaced to the right for the sake of clarity). Galactic and extragalactic cases (as defined above) are shown in the left and right panels, respectively. The effect of a $\sim 0.3$ Jy photometric error on the flux density of the sources is negligible in our simulated experimental setting. This comes not as a suprise, since a $\sim 0.3$ Jy variation in $S_0$ produces only a $\sim 10 \%$ change in the $\mu_1$ term that appears in (\ref{eq:estimator}) and (\ref{eq:estimatorb}) in the worst case (1 Jy sources)\footnote{Moreover, the estimation of the polarization is not given directly by (\ref{eq:estimator}) and (\ref{eq:estimatorb}), but by the minimization of these functions. A small variation in one of the terms of the functions does not necessarily mean that the \emph{position} of the minimum of the function changes in a noticeable way. The non linear way in which $S_0$ appears in these equations makes it difficult to find an analytical expression of how an uncertainty in $S_0$ affects the minimization. This question is better answered by simulations, just as we have done in this section.}. This discrepancy quickly decreases as $S_0$ grows. Moreover, the rms around the mean $\langle \mu_1 \rangle$ also decreases very quickly with $S_0$. Therefore, we conclude that our Bayesian estimator is robust against moderate uncertainties on the prior and the flux density of the sources.

\section{Conclusions} \label{sec:conclusions}

The estimation of the polarimetric properties of extragalactic compact sources at microwave wavelengths will be very relevant in the upcoming years. In this work, we have introduced a Bayesian approach for the estimation of the polarized flux density $P$ of this kind of sources. Following recent works by \cite{massardi13,galluzzi17,galluzzi19} among others, we have proposed an analytical prior for the polarization fraction of extragalactic radio sources which takes the form of a log-normal distribution whose parameters (median, average and variance values of the polarization fraction) can be constrained by the latest observational data. Using this prior, we have proposed two maximum \emph{a posteriori} (MAP) estimators of the polarization of a given source given observations of its $Q$ and $U$ Stokes parameters. The first method works directly on the quadratic combination $P^2 = Q^2 + U^2$ whereas the second method produces individual estimators of the ground-truth values $Q_0$ and $U_0$ that are then quadratically added to give an estimator of the ground-truth polarization $P_0$ of the source.  We have called these methods Bayesian Rice and Bayesian Filtered Fusion (BFF), respectively. Both can be considered as natural Bayesian extensions of the frequentist Neyman-Pearson and standard Filtered Fusion (FF) methods introduced by \cite{argueso09}.
The standard FF is shown to be equal to the Maximum Likelihood Estimator (MLE) for $P$, whereas the BFF adds 
to the MLE a number of additional terms that include the \emph{a priori} information on the distribution of the polarization fraction.
The BFF method can be easily accommodated to non-white noise and foregrounds. For this reason we have focused on this method in most of our paper.

We have tested the performance of the BFF method and compared it to that of FF using two sets of simulations: polarized sources embedded in $Q$ and $U$ white noise, and more realistic simulations that include also polarized CMB and Galactic foreground emission. In both cases we have used the pixel and beam scales plus the noise levels and sky coverage of the QUIJOTE experiment Wide Survey \citep{mfiwidesurvey,quijote_PS} at 11 GHz. 
For the BFF we assumed that the flux density $S_0$ of the sources is perfectly known.
For highly polarized sources the two methods yield the same results, but for medium to low polarizations ($P_0 \leq 400$ mJy in our simulations) the BFF gets more accurate estimations of the polarization of the sources. The FF
 gets noise-limited around a polarization flux $P_0 \sim 500$ mJy, whereas the BFF allows us to reach polarized fluxes well below $P_0 \sim 100$ mJy before becoming noise-limited itself. 
 Both estimators are biased for low polarization (i.e. $P_0 \lesssim 500$ mJy) sources: the BFF
 tends to underestimate the polarization, whereas the standard FF overestimates the polarization of these sources. In the case of the FF the bias is due to noise boosting of the signal (akin to Eddington bias). In the case of the BFF, the bias is originated by the extra terms in the estimator formula that come from the physical prior. However, the absolute value of bias is significantly smaller for the BFF than for the FF, specially for faint sources. 
 
In the above discussion, we have assumed that the prior describes the real distribution of polarization of the sources adequately and that the total flux density $S_0$ of each source is perfectly known. However, information about the polarization properties of extragalactic sources at microwave frequencies above $\simeq 10$ GHz is still scarce. Moreover, for any given source $S_0$ is known with a certain degree of uncertainty (due to instrumental noise, less-than-perfect modelling of the spectral energy distribution of the source and variability, among other possible causes).  In the last part of this work, we have tested the robustness of the BFF estimator against moderate changes in the prior parameters and realistic uncertainties in the flux density of the sources. Our simulations indicate that assuming the wrong prior has a mild effect on the Bayesian estimator. For example, for a $S_0=10$ Jy source, a change by a factor of two in the assumed median polarization fraction of the sources introduces errors or the order $\lesssim 100$ mJy in the estimation of $P$. Regarding uncertainties in the flux density of the sources, we find that non-catastrophic photometric error bars have a minimal impact on the estimation of $P$.

 We  therefore conclude that the Bayesian approach can significantly improve the estimation of the polarization of extragalactic radio sources in current and upcoming CMB polarization experiments.  In an upcoming work, we will explore the extension of the Bayesian framework to the multi-frequency case.

\begin{acknowledgements}
 We thank the Spanish MINECO 
 and
the Spanish Ministerio de Ciencia, Innovaci\'on y Universidades for
 for partial financial
support under projects AYA2015-64508-P  and
PGC2018-101814-B-I00, respectively. DH also acknowledges 
funding from the European Union’s Horizon 2020 research and innovation
programme (COMPET-05-2015) under grant agreement number 687312 (RADIOFOREGROUNDS). 
Some of the results in this paper have been derived using the  HEALPix \citep{healpix} and \texttt{healpy} \citep{Zonca2019} packages. This research made use of \texttt{astropy},\footnote{http://www.astropy.org} a community-developed core Python package for Astronomy \citep{astropy:2013, astropy:2018},  \texttt{matplotlib}, a Python library for publication quality graphics \citep{Hunter:2007}, and \texttt{SciPy}, a Python-based ecosystem of open-source software for mathematics, science, and engineering \citep{Virtanen_2020}.
We acknowledge Santander Supercomputacion support group at the University of Cantabria (UC) who provided access to the supercomputer Altamira Supercomputer at the Institute of Physics of Cantabria (IFCA-UC-CSIC), member of the Spanish Supercomputing Network\footnote{\url{https://www.res.es/en/about}}, for performing simulations/analyses. 
\end{acknowledgements}

\bibliographystyle{aa} % style aa.bst
\bibliography{FFbayes.bib,planck_bib.bib} % your references Yourfile.bib

\label{lastpage}

\end{document}